\documentclass[12pt,a4paper]{article}
\usepackage{amssymb}
\usepackage{amsmath,amsfonts,amsthm}

\usepackage{graphicx}
\usepackage{amsmath}

\usepackage{amsfonts}
\usepackage{amsthm,amscd}
\usepackage{graphicx}
\usepackage{amsmath}
\usepackage{amssymb}
\usepackage{amstext}

\newtheorem{theorem}{Theorem}
\newtheorem{corollary}[theorem]{Corollary}
\newtheorem{definition}[theorem]{Definition}

\newtheorem{lemma}[theorem]{Lemma}

\title{ A detailed  proof  of the von Neumann's Quantum Ergodic Theorem}

\author{Artur O. Lopes \, and Marcos Sebastiani}

\begin{document}

\maketitle

\begin{abstract}

We present a simplified proof of the von Neumann's Quantum Ergodic Theorem. This important result was initially published in german by J. von Neumann in 1929.

We are interested here in the time evolution $\psi_t$, $t\geq 0$, (for large times) under the Schrodinger equation
associated to a given fixed Hamiltonian $H : \mathcal{H} \to \mathcal{H}$ and a general initial condition $\psi_0$.
The  dimension of the Hilbert space  $\mathcal{H}$ is finite.

\end{abstract}

\section{Introduction}

Consider a fixed Hamiltonian $H$ (a complex self adjoint operator) acting on a complex Hilbert space $\mathcal{H}$ of dimension $D$, where $D\geq 3$. Then, $\mathcal{H}$ can be written as
$$\mathcal{H}\,=\, \mathcal{V}_1\, \oplus ...\oplus \mathcal{V}_K,$$
where each $\mathcal{V}_a$, $a=1,2,...,K$, is the subspace of eigenvectors associated to the eigenvalue $\lambda_a$,
and
$\lambda_1< \lambda_2<...<\lambda_K.$

%A general reference on Quantum Mechanics for mathematicians appears in \cite{Lop}.

We fixed an initial condition $\psi_0$ for the dynamic Schrodinger evolution.
We consider the time evolution $\psi_t = e^{-i\,t\,H} (\psi_0)$, $t \geq 0$, and we are interested in properties for most of the large times (not all large times).

Now we consider another decomposition $\mathcal{D}$  of $\mathcal{H}$ (which has nothing to do with the previous one)
$$\mathcal{H}\,=\, \mathcal{H}_1\, \oplus ...\oplus \mathcal{H}_N,\,\,\,N\geq 2.$$

We can consider a natural probability on the set $\Delta$ of possible decompositions $\mathcal{D}$ and we are interested here in properties for most of the decompositions $\mathcal{D}$. For small $\delta>0$ we are interested in the concept of a $(1-\delta)$ generic decomposition $\mathcal{D}$ (in the probabilistic sense).

For a given fixed subspace   $\mathcal{H}_\nu$  of $\mathcal{H}$, $\nu=1,...,N$, the observable $P_{\mathcal{H}_\nu  }$ (the orthogonal projection on $\mathcal{H}_\nu$)
is such that the mean value of the state $\psi_t$, $t \geq 0$, is given by  $E_{\psi_t} (P_{\mathcal{H}_\nu}) =<P_{\mathcal{H}_\nu  }(\psi_t), \psi_t>= |P_{\mathcal{H}_\nu  }(\psi_t)\,|^2.$

In the first part of the paper, following the basic guidelines of the original work by J. von Neumann, we present lower bound conditions (in terms of $\delta$, etc...) on the dimensions $d_\nu$, $\nu=1,2,..,N$, of the different
$\mathcal{H}_\nu $ of a $(1-\delta)$-generic orthogonal decomposition $\mathcal{D}$ of the form $\mathcal{H}\,=\, \mathcal{H}_1\, \oplus ...\oplus \mathcal{H}_N$, in such way that the dynamic time evolution $\psi_t$, $t \geq 0$, of a given $\psi_0$,
for most of the large times $t$, has the property that the expected value $E_{\psi_t} (P_{\mathcal{H}_\nu}) $ is almost $\frac{d_\nu}{D}$. In this way there is an approximately  uniform spreading of $\psi_t$ among the different $\mathcal{H}_\nu$ of a generic  decomposition $\mathcal{D}$. In this part the main result is Theorem \ref{t1}. We point out that these estimates are for a fixed initial condition $\psi_0$.

The  von Neumann's Quantum Ergodic Theorem provides uniform estimates for all $\psi_0$. This result is presented in Theorem \ref{main}.
This will be done in the second part of the paper which begins on section \ref{uni}.
In order to get this theorem it will be necessary to assume hypothesis on the eigenvalues of the Hamiltonian $H$ (see hypothesis $\mathfrak{N\,\,R}$ just after Lemma \ref{a9}).
\medskip

Suppose for instance that $A: \mathcal{H} \to \mathcal{H}$ is an observable and this self adjoint operator has spectral decomposition
$$\mathcal{H}\,=\, \mathcal{H}_1\, \oplus ...\oplus \mathcal{H}_N,$$
where $  \mathcal{H}_p$, $p=1,...,N$ is the subspace of eigenvectors associated to the eigenvalue $\beta_p$, and
$\beta_1< \beta_2<...<\beta_N.$ The probability that the measurement of $A$ on the state $\psi_t$ is $\beta_p$ is given by
$<P_{\mathcal{H}_p  }(\psi_t), \psi_t>$. This shows the relevance of the result. The point of view here is not to look for generic observables but for generic decompositions.
\medskip

We stress a point raised on \cite{Leb}. What is proved is a property of the kind:  for  most $\mathcal{D}$ something is true for all $\psi_0$. And, not a property of the kind: for all $\psi_0$, something is true for most $\mathcal{D}.$

Of course, the main result can also be stated in terms of limits, when $T\to \infty$, of means $ \frac{1}{T} \int E_{\psi_t} (P_{\mathcal{H}_\nu}) dt$, which is a more close expression to the one present in the classical Ergodic Theorem.

\bigskip

We  present here a simplified proof (with less hypothesis in some parts) when dim $\mathcal{H}$ is finite of this important result which was initially published in german by J. von Neumann in 1929 (see \cite{von}).  The paper \cite{Tu} presents a translation from german to english of this work of  von Neumann. This 1929 paper also consider the concept of Entropy for such setting. We will not consider this topic in our  note.

Several papers with interesting discussions about this work appeared recently (see for instance \cite{Tu}, \cite{Leb} and other papers which mention these two)

\medskip

Consider a general connected compact Riemannian manifold  $X$  and its volume form. When  properly normalized  this procedure defines a natural probability $w_X$  over $X$.

Given  a compact  Lie group (real) $G$, one can consider the associated bi-invariant Riemannian metric. If $H$ is a closed subset of $G$, this metric can be considered in the quotient space $X= \frac{G}{H}$ and in this way we get a probability on such manifold $X$. We will denote by $\pi$ the projection.

When we consider expected values  of a function $f$ this we will be taken with respect to the above mentioned probability.

\begin{lemma} \label{Lem1}

Given a continuous function $f:X \to \mathbb{C}$ and $\pi: G\to X$ the canonical projection, then
$$a) \,\,\text{vol}\, (S) =   \frac{\text{vol}\, (\,\pi^{-1} (S)\,)} {\text{vol}\, (H) } $$
for every Borel set $S\subset X$, and

$$ b)\,\, E_X(f)= E_G (f \circ \pi).$$

\end{lemma}

The first integral is taken with respect to the volume form $w_X$ and the second with respect to the volume form $w_G$.

Note that vol $(G)=$ vol $(X)\,$ vol $(H).$

The proof is left for the reader.

\medskip

Suppose $\mathcal{H}$ is a complex Hilbert space of finite dimension $D$ with an inner product $<\,,\,>$ and a norm $|\,\,\,|$.

Suppose we fix a decomposition $ \mathcal{D}$, that is,
$$ \mathcal{D}\,:\, \mathcal{H}\,=\, \mathcal{H}_1\, \oplus ...\oplus \mathcal{H}_N$$
$ N>1$, is a orthogonal direct sum where  dim $\mathcal{H}_\nu=d_\nu>0$ for all $\nu=1,2,...,N.$

Denote $P_\nu$ the orthogonal projection of $  \mathcal{H}\,$ over \, $\mathcal{H}_\nu$.

Moreover, $S=\{ \psi \in  \mathcal{H}\,|\, | \psi|=1\} $ denotes the unitary sphere. $S$ has a Riemannian structure with  a metric induced by the norm in $\mathcal{H}.$ In the same way as before, there is an associated probability $w_S$ is $S$.

\begin{lemma} \label{a0} For any $\nu=1,2...,N$,

$$\, E_S(\,| P_\nu\, (\,.\,)\,|^2)=   \int_S\, | P_\nu\, (\phi)\,|^2\, d\, w_S (\phi)\,=\frac{d_\nu}{D} .$$

\end{lemma}

{\bf Proof:}

Suppose $\nu$ is fixed, then take $\psi_1,\psi_2,...,\psi_D$, and orthogonal basis of $\mathcal{H}$, such that, $\psi_1,\psi_2,...,\psi_{d_\nu}$ is an orthogonal basis of $\mathcal{H}_\nu.$

Given $\phi= \sum_{j=1}^D  x_j\, \psi_j \in S,$ where $\sum_{j=1}^D  |x_j|^2=1$, then
$$ \int_S\, | P_\nu\, (\phi)\,|^2\, d\, w_S (\phi)= \int_S \,\sum_{j=1}^{d_\nu}  |x_j|^2 d\, w_S (x).
$$

Note that the integral $\int_S \, |x_j|^2 d\, w_S (x)$ is independent of $j$ and
$$\int_S \,\sum_{j=1}^{D}  |x_j|^2 d\, w_S (x) = \,\text{vol}\,\,(S)=1.$$

Therefore, for any $j$
$$\int_S \, |x_j|^2 d\, w_S (x)=\,\frac{1}{D}\, .$$

Therefore, it follows that
$$\int_S \,\sum_{j=1}^{d_\nu}  |x_j|^2 d\, w_S (x)=\,\frac{d_\nu}{D}\, .
$$

\qed

\medskip

\begin{lemma} \label{a1} For any $\nu=1,2...,N$,

$$\, \text{Var}_S(\,| P_\nu\, (\,.\,)\,|^2)=   \int_S\, (\,| P_\nu\, (\phi)|^2\,- \frac{d_\nu}{D})^2\, d\, w_S (\phi)\,=\frac{d_\nu\, (D- d_\nu)}{D^2 \, ( D+1 )} .$$

\end{lemma}

{\bf Proof:} In order to simplify the notation we take $\nu=1$. Then, we denote $d=d_1$ and $P=P_1$.

Take $\psi_1,\psi_2,...,\psi_D$, and orthogonal basis of $\mathcal{H}$, such that, $\psi_1,\psi_2,...,\psi_{d}$ is an orthogonal basis of $\mathcal{H}_1.$

By last Lemma we have
$$ \int_S\,(\, | P\, (\phi)|^2\,- \frac{d}{D})^2\, d\, w_S (\phi)\,= $$
$$\int_S\, | P\, (\phi)|^4\, d\, w_S (\phi)\,- 2\,\frac{d}{D} \, \int_S\, | P\, (\phi)|^2\, d\, w_S (\phi)+ (\frac{d}{D})^2=$$
$$\int_S\, | P\, (\phi)|^4\, d\, w_S (\phi)\,- (\frac{d}{D})^2.$$

If $\phi= \sum_{j=1}^D  x_j\, \psi_j \in S,$ then $P(\phi) =\sum_{j=1}^d  x_j\, \psi_j .$

Therefore,

$$ \int_S\, | P\, (\phi)|^4\, d\, w_S (\phi)\,=\,\frac{1}{\text{vol (S)}   } \int_S \, (\,\sum_{j=1}^d  |x_j|^2\,)^2\, d\, S(x)= \, \frac{d^2 + d}{D\, (D+1)}.$$

The last equality follows from a standard computation (see Appendix 1).

From this follows the claim.

\qed

\medskip

\bigskip

\section{Changing the decomposition} \label{sec2}

$\mathcal{H}$ is fixed for the rest of the paper.

Now we change our point of view. We fix $\phi \in \mathcal{H}$ and we consider different decompositions of $\mathcal{H}$ in direct sum. More precisely, we fix
$D=$ dim $\mathcal{H}$ and $N$ and we consider fixed natural positive numbers $d_\nu$, $\nu=1,2,...,N$, such that $d_1+d_2+...+d_N=D$, and then, all possible choices of orthogonal decompositions with this data.

We denote by $\Delta(d_1,d_2,...,d_N, \mathcal{H} ) = \Delta$ the set of all possible $ \mathcal{D}$, that is, all possible orthogonal direct  sum decompositions
$$ \mathcal{D}\,:\, \mathcal{H}\,=\, \mathcal{H}_1\, \oplus ...\oplus \mathcal{H}_N.$$

For fixed $\nu=1,2,...,N$, then $ P_\nu (\mathcal{D})$ denotes the projection on $\mathcal{H}_\nu$ associated to the decomposition $\mathcal{D}$.

Each choice of  orthogonal basis $\psi_1,\psi_2,...,\psi_D$  of $\mathcal{H}$, defines a possible choice of direct orthogonal sum decomposition:
$$ \mathcal{H}_1 \, \,\text{ is generated by}\,\,\{\psi_1,...,\psi_{d_1}\,\},\,\,\mathcal{H}_2 \, \,\text{ is generated by}\,\,\{\psi_{d_1+1},...,\psi_{d_1+ d_2}\,\}\,,$$
and, so on.

The set of all orthogonal basis is identified with the set of unitary operators $U(D)$ which defines a compact Lie group and a Haar probability structure.

In this way,
$$ \Delta= \frac{U(D)}{U(d_1) \times U(d_2)\times...\times U(d_N)}.$$

In the same way as before we get a probability $w_\Delta$ over $\Delta$. Therefore, it has a meaning the probability $w_\Delta(B)$ of a Borel set $B\subset \Delta$ of decompositions.

\begin{lemma} \label{a2} Consider a continuous function $f: \mathbb{R} \to \mathbb{R}$. Then, for fixed $\nu=1,2...,N$, and fixed
$\tilde{\phi}$ and  $\tilde{\mathcal{D}}$

$$\,    \int_S\, f(\,| P_\nu (\tilde{\mathcal{D}}) \, \phi\,|\,)\,d\, w_S (\phi)\,=   \int_\Delta\, f(\,| P_\nu (\mathcal{D}) \, \tilde{\phi}\,|\,)\,d\, w_\Delta (\mathcal{D}).$$

This constant value is independent  of  $\tilde{\phi}$ and  $\tilde{\mathcal{D}}$.

\end{lemma}

{\bf Proof:} If $U: \mathcal{H} \to \mathcal{H},  $ is unitary, then $U\, \mathcal{D}$ denotes
$$ U(\mathcal{H}_1)\, \oplus ...\oplus U(\mathcal{H}_N).$$

\medskip

Then, for fixed $\phi$ and $\mathcal{D}$ we have
$$ P_\nu (U\, \mathcal{D})\, U \,(\,\phi \,)= \, U \,P_\nu ( \mathcal{D}) \phi.  $$

We prove the claim for $P_1$. Suppose $\psi_1,\psi_2,...,\psi_D$, is an orthogonal basis of $\mathcal{H}$, such that, $\psi_1,\psi_2,...,\psi_{d_1}$ is an orthogonal basis of $\mathcal{H}_1.$

We can express $\phi= \sum_{j=1}^D x_j \, \psi_j$ and moreover $U(\phi)= \sum_{j=1}^D x_j \, U( \psi_j)$.

$U(\psi_1),U(\psi_2),...,U(\psi_D)$ is an orthogonal basis of $\mathcal{H}$ associated to $U\, \mathcal{D}$ and
$U(\psi_1),U(\psi_2),...,U(\psi_{d_1})$ is an orthogonal basis of $U(\mathcal{H}_1)$.

Then,
$$ P_1 (U\, \mathcal{D})\, U \,(\,\phi \,)= P_1 (U\, \mathcal{D})\, (\sum_{j=1}^D x_j \, U( \psi_j))=\sum_{j=1}^{d_1} x_j \, U( \psi_j).$$

By the other hand
$$ U \,P_1 ( \mathcal{D}) \phi =U \,P_1 ( \mathcal{D}) (\sum_{j=1}^D x_j \, \psi_j\,)= U (  \sum_{j=1}^{d_1} x_j \, \psi_j) =\sum_{j=1}^{d_1} x_j \, U( \psi_j), $$
and this shows the claim.

Therefore, we get
$$  |\,P_\nu (U\, \mathcal{D})\, U \,(\,\phi \,)\,|\, = |\,U^{-1}\,P_\nu (U\, \mathcal{D})\, U \,(\,\phi \,)\,|\,= |\,U^{-1}  U \,P_\nu ( \mathcal{D}) \phi| = |\,P_\nu ( \mathcal{D}) \phi|.$$

Finally, for a fixed $\mathcal{D}$ and a variable $U$

$$\,    \int_S\, f(\,| P_\nu (\mathcal{D}) \, \phi\,|\,)\,d\, w_S (\phi)\,=  \int_S\, f(
|\,P_\nu (U\, \mathcal{D})\, U \,(\,\phi \,)\,|\,)\,d\, w_S (\phi)\,= $$
$$  \int_S\, f(
|\,P_\nu (U\, \mathcal{D})\, \,(\,\phi \,)\,|\,)\,d\, w_S (\phi)\,,$$

because $w_S$ is invariant by the action of $U$.

Then the above integral on the variable $\phi$ is constant by the action of $U$ in  a given  decomposition $\mathcal{D}$.
\medskip

Now consider a fixed $\phi_1$ and another general  $\phi_2 = U ( \phi_1)$, where $U$ is unitary.

As $w_\Delta$ is  invariant by the action of $U$ the integral
$$
\int_\Delta\, f(\,| P_\nu (\mathcal{D}) \, \phi_2\,|\,)\,d\, w_\Delta (\mathcal{D})= \int_\Delta\, f(\,| P_\nu ( U\, \mathcal{D}) \, U ( \phi_1) \,|\,)\,d\, w_\Delta ( \mathcal{D})=$$
$$\int_\Delta\, f(\,| U \, \,P_\nu (\mathcal{D}) \, \phi_1\,|\,)\,d\, w_\Delta (\mathcal{D})= \int_\Delta\, f(\,| \, \,P_\nu (\mathcal{D}) \, \phi_1\,|\,)\,d\, w_\Delta (\mathcal{D})$$
is constant and independent of $\phi$.

Remember that $ w_S \times w_\Delta$ is a probability.

Consider now

$$ \int \int   f(\,| P_\nu (\mathcal{D}) \, \phi\,|\,)\,d\, w_S (\phi) d\, w_\Delta (\mathcal{D})= $$
$$ \int\, [\,\, \int   f(\,| P_\nu (\mathcal{D}) \, \phi\,|\,)\,d\, w_S (\phi)\,\,]\, \,d\, w_\Delta (\mathcal{D})= $$
$$ \int \,[\,\,\int   f(\,| P_\nu (\mathcal{D}) \, \phi\,|\,)\,d\,w_\Delta (\mathcal{D})\,\,]\,\,d\, w_S (\phi) ,  $$

then by Fubini we get the claim of the Lemma (since the unitary group acts transitively on $S$ and on $\Delta$).
\medskip

\qed

\medskip

\begin{corollary} \label{co1} Consider a fixed $\phi \in \mathcal{H}$ such that $|\phi|=1$.

Then, for $\nu=1,2...,N$, we get that

$$\, E_\Delta(\,| P_\nu\,(  \,.\,) (\,\phi\,)\,|^2)=   \frac{d_\nu}{D} ,$$
and
$$\, \text{Var}_\Delta(\,| P_\nu\,(  \,.\,) (\,\phi\,)\,|^2)=   \frac{d_\nu\, (D- d_\nu)}{D^2 \, ( D+1 )} ,$$
where $\,.\,$ denotes integration with respect to $\mathcal{D}.$

\end{corollary}

{\bf Proof:} This is consequence of Lemmas \ref{a0}, \ref{a1} and \ref{a2}.

\qed

\medskip

\begin{definition} Given $\delta>0$, a Hilbert space $\mathcal{H}$ and natural positive numbers $d_j, j=1,2,...,N$, such that, $d_1 + d_2 +...+ d_N=D=$ dim $\mathcal{H}$, we say that a property is true for $\mathcal{D}\in \Delta (d_1,..,d_N, \mathcal{H})$, in $(1-\delta)$ sense, if the property is not true only for  elements $\mathcal{D}$ in a set of probability $w_\Delta$ smaller than  $\delta$.

\end{definition}

\medskip

\begin{corollary} \label{gos}  Suppose $\epsilon>0$ and $\delta>0$ are given. Consider natural positive numbers $d_\nu, \nu=1,2,...,N$, such that, $d_1 + d_2 +...+ d_N=D=$ dim $\mathcal{H}$, and moreover assume that, for all $\nu=1,2...,N$,
$$ d_\nu > D - \frac{\epsilon^2\, \delta D\, (D+1) }{N^2}.$$

Consider a fixed $ \phi$ such that $|\phi|=1$. Then, for decompositions  $\mathcal{D}\in\Delta (d_1,..,d_N, \mathcal{H})$ in the  $(1-\delta)$ sense, and $\nu=1,2...,N$, we have
\begin{equation} \label{esta}
|\,\,| P_\nu\,(  \,\mathcal{D}\,) (\,\phi\,)\,|^2\,-\,\frac{d_\nu}{D}\,\,\,| < \epsilon \sqrt{\frac{d_\nu}{D\,N} } .
\end{equation}

\end{corollary}

{\bf Proof:} By Corollary \ref{co1} and Markov inequality  we have
$$ w_\Delta ( \,[\,| P_\nu\,(  \,\mathcal{D}\,) (\,\phi\,)\,|^2\,-\,\frac{d_\nu}{D}\,\,]^2\,\geq \epsilon^2 \frac{d_\nu}{D\,N} \,\, )\,\leq $$
$$    \frac{d_\nu\, (D- d_\nu)}{D^2 \, ( D+1 )} \, \frac{D\,N}{\epsilon^2\,d_\nu}=  \frac{N\, (D- d_\nu)}{\epsilon^2\, D \, ( D+1 )}. $$

Then, the probability that all $N$ inequalities do not happen is

$$ 1- N\,  \frac{N\, (D- d_\nu)}{\epsilon^2\, D \, ( D+1 )}>1- \delta $$
by hypothesis.

\qed

\medskip

The corollary above means that for a fixed $\phi$, if the $d_\nu$ are all not very small, then for a big part of the decompositions
$ \mathcal{D}$ we have that
$$ \,\,| P_\nu\,(  \,\mathcal{D}\,) (\,\phi\,)\,|^2\,$$
is close by the mean value $\,\frac{d_\nu}{D}$.

\bigskip

\begin{definition} Given a Hilbert space $\mathcal{H}$ and  a fixed decomposition $\mathcal{D}$ (associated to natural positive numbers $d_j, j=1,2,...,N$, such that, $d_1 + d_2 +...+ d_N=D=$ dim $\mathcal{H}$, we define a semi-norm in such way that  for a linear operator $\rho :\mathcal{H} \to \mathcal{H}, $ by
$$ |\,\rho\,|_\infty = |\,\rho\,|_\infty^\mathcal{D}= \sup_{1 \leq \nu \leq N} \, |\, \text{Tr} \, (\rho\,\, P_\nu (\mathcal{D}  )\,|  $$

\end{definition}

\medskip

The above means that if $|\,\rho\,|_\infty$ is small, then all expected values $E_{P_\nu} (\rho) $, $\nu=1,2,...,N$,  are small
\medskip

$|\,\phi>\,<\phi\,|$ will denote the orthogonal projection on the unitary vector $\phi$ in the Hilbert space $\mathcal{H}$.

\medskip

\begin{lemma} \label{a3} Consider a  $\phi \in  \mathcal{H}=\mathcal{H}_1\, \oplus ...\oplus \mathcal{H}_N$ such that $ |\phi|=1$.  Denote $\rho_{mc} = \frac{1}{D} I_{\mathcal{H}}.$

Then,

$$ |\, \, |\,\phi>\,<\phi\,|\, -\, \rho_{mc}\,|_\infty\,\,= \sup_{ 1 \leq \nu \leq N }\, |\,\,|  P_\nu (\mathcal{D})\,(\phi) \,|^2 - \frac{d_\nu}{D}\,\,|.$$

\end{lemma}

{\bf Proof:}

Suppose  $\psi_1,\psi_2,...,\psi_D$ is orthogonal basis of $\mathcal{H}$, such that, $\psi_1,\psi_2,...,\psi_{d_1}$ is an orthogonal basis of $\mathcal{H}_1.$

If $\phi = \sum_{j=1}^D x_j \phi_j$, then, for $i=1,2,...,d_1$,

$$  |\,\phi>\,<\phi\,|\,| P_1 (\phi_i)>= |\,\phi>\,<\phi\,|\,|\phi_i>\,= \sum_{j=1}^D \,\overline{x_i}\,x_j\, \phi_j
$$
and
$$  |\,\phi>\,<\phi\,|\,| P_1 (\phi_i)>=0$$
for $i> d_1$.

Therefore,
$$ \text{Tr}\,\,[\, |\,\phi>\,<\phi\,|\,| P_1\,(.)\,>\,] = \sum_{j=1}^{d_1} \,\,|x_j|^2= \, | P_1 (\phi)|^2.$$

In an analogous way we have that for any $\nu$
$$ \text{Tr}\,\,[\, |\,\phi>\,<\phi\,|\,| P_\nu\,(.)\,>\,]= \, | P_\nu (\phi)|^2.$$

From this follows the claim.

\qed

\medskip

From the above it follows:

\begin{corollary} Under the hypothesis of Corollary \ref {gos}, we get that for decompositions  $\mathcal{D}\in\Delta (d_1,..,d_N, \mathcal{H})$ in the  $(1-\delta)$ sense,

$$ |\, \, |\,\phi>\,<\phi\,|\, -\, \rho_{mc}\,|_\infty\,\,\leq  \sup_{ 1 \leq \nu \leq N }\, \epsilon\, \sqrt{\frac{d_\nu}{N\, D}}.$$

\end{corollary}

\qed

\medskip

\bigskip

\section{Estimations on time}

\begin{definition} Given $\delta>0$ we say that a property for the parameters $t\in \mathbb{R}$  is true for  $(1-\delta)$-most of the large times, if
$$ \liminf_{T \to \infty} \, \frac{1}{T} \mu (A_T)\,\,> \,\,1- \delta,$$
where $A_T$ is the set of $t\in [0,T]$ where the property is verified and $\mu$ is the Lebesge measure on $\mathbb{R}.$

\end{definition}

\medskip

\begin{lemma} \label{a5} Suppose $f: \mathbb{R} \to \mathbb{R}$ is continuous and non negative. Consider a certain $\gamma>0$.

Suppose $\rho$ is such that
 $$ \limsup_{T \to \infty} \, \frac{1}{T} \int_0^T f(t)\,\, dt\,<\, \rho.$$

Then, $f(t)< \gamma$ for $1-\frac{\rho}{\gamma}$-most of the large times.

\end{lemma}

{\bf Proof:}

$$ \int_0^T f(t)\, dt \geq \int_{f(t)\geq \gamma}^T f(t) \, dt \geq \gamma\, \mu ( \{ t\in [0,T] \,|\, f(t)\geq \gamma  \}\,).$$

Therefore,
$$ \limsup_{T \to \infty} \, \frac{1}{T} \mu( \, \{\, t \in [0,T] \, f(t) \geq  \gamma\,\}\, \,<\, \frac{\rho}{\gamma},$$
and finally

$$ \liminf_{T \to \infty} \, \frac{1}{T} \mu( \, \{\, t \in [0,T] \, f(t)< \gamma\,\}\, \,>\, 1-\frac{\rho}{\gamma}.$$

\qed

\medskip

Suppose $\mathcal{H}$ is  Hilbert space, $d_j, j=1,2,...,N$, are such that, $d_1 + d_2 +...+ d_N=D=$ dim $\mathcal{H}$, and $H : \mathcal{H} \to \mathcal{H}$ a selfadjoint operator.
Consider a fixed $\phi_0 \in \mathcal{H}$, with $|\phi_0|=1$, and $\psi_t = e^{-\,i\, t \, H}\, \phi_0$, $t\geq 0$, a solution of the associated Schrodinger equation.

\medskip

\begin{lemma} \label{a6} For fixed $T$ and $\nu=1,2,...,N$, consider the function
$$f_{\nu,T} : \Delta(d_1,d_2,...,d_N, \mathcal{H} ) \times \, S\to \mathbb{R},$$
given by
$$f_{\nu,T} (\mathcal{D}, \phi)\,\,=\,\, \frac{1}{T}\, \int_0 ^T \, (\, | P_\nu (\mathcal{D})\, \psi_t\,|^2 -\frac{d_\nu}{D}   )^2 \, dt.$$

Then, $f_{\nu,T}$
converges uniformly on $(\mathcal{D},\phi) \in \Delta(d_1,d_2,...,d_N, \mathcal{H} )\times S$, when $T\to \infty$, for any $\nu=1,2,...,N$.

\end{lemma}

{\bf Proof:}

Suppose $\phi_1,\phi_2,...,\phi_D$ is a set of eigenvectors of $H$ which is an orthonormal basis of $\mathcal{H}$.

Assume that $\phi_0= \sum_{j=1}^D x_j \phi_j$. Then
$$ \psi_t = \sum_{j=1}^D \, x_j \, e^{ -i\,t\, E_j}\, \phi_j, $$
where $E_j$, $j=1,2,..,D$ are the corresponding eigenvalues.

Then, for a given $\nu$
$$| P_\nu (\mathcal{D})\, \psi_t\,|^2= < \psi_t, P_\nu (\mathcal{D})\,(\psi_t)> = \sum_{\alpha,\beta}  x_\alpha \overline{x_{\beta}}  e^{ -i\,t\, (E_\alpha- E_\beta)}\phi_j< \phi_\alpha, P_\nu (\mathcal{D})\,(\phi_\beta)> .$$

Therefore,

$$ (\, | P_\nu (\mathcal{D})\, \psi_t\,|^2 -\frac{d_\nu}{D}   )^2 \,=\, \sum_{w=1}^M\, L_{w,\nu} ( \mathcal{D},\phi) \,e^{i\, u_w t},$$

where $M \in \mathbb{N}$, $u_1,..,u_M$ are real constants and $ |\,L_{w,\nu}(\mathcal{D},\phi) \,|\leq 2$.

Then,
$$ f_{\nu,T} (\mathcal{D},\phi)=\sum_{u_w=0}^M\ L_{w,\nu}(\mathcal{D},\phi)+  \frac{1}{T} \sum_{u_w\neq 0}^M\ L_{w,\nu}(\mathcal{D},\phi)(\, \frac{e^{i\,u_w\, T}}{i\, u_w} - \frac{1}{i \,u_w}\,) .$$
Finally, we get

$$ |\, f_{\nu,T} (\mathcal{D},\phi) - \sum_{u_w=0}^M\ L_{w,\nu}(\mathcal{D},\phi)\,| \leq \frac{1}{T}\, \frac{4\,M}{\inf_{u_w\neq 0} \, |u_w|}  .$$

As $M$ is fixed the claim follows from this.

\qed

\medskip

\begin{corollary}
\label{a8}
$$ \int_\Delta\,(\,\,\lim_{T \to \infty}  \frac{1}{T}\, \int_0 ^T \, (\, | P_\nu (\mathcal{D})\, \psi_t\,|^2 -\frac{d_\nu}{D}   )^2 \, dt\,\,)\,\, d w_\Delta (\mathcal D)\,=   \frac{d_\nu\, (D- d_\nu)}{D^2 \, ( D+1 )} ,$$
for any $\nu=1,2,..,N$.
\end{corollary}

{\bf Proof:}
By Lemma \ref{a6} and Corollary \ref{co1}  we have that
$$ \int_\Delta\,[\,\,\lim_{T \to \infty}  \frac{1}{T}\, \int_0 ^T \, (\, | P_\nu (\mathcal{D})\, \psi_t\,|^2 -\frac{d_\nu}{D}   )^2 \, dt\,\,]\,\, d w_\Delta (\mathcal D)\,=$$
$$\lim_{T \to \infty}  \frac{1}{T}\,\int_\Delta\, d w_\Delta (\mathcal D)\,\,(\,\, \int_0 ^T \, (\, | P_\nu (\mathcal{D})\, \psi_t\,|^2 -\frac{d_\nu}{D}   )^2 \, dt\,\,) $$
$$ \lim_{T \to \infty}  \frac{1}{T}\,\int_0^T\, dt\, \int_\Delta\,\, (\, | P_\nu (\mathcal{D})\, \psi_t\,|^2 -\frac{d_\nu}{D}   )^2 \,d w_\Delta (\mathcal D)\, = \frac{d_\nu\, (D- d_\nu)}{D^2 \, ( D+1 )}.$$
\qed

\medskip

\begin{theorem}
\label{t1}

Suppose $\epsilon>0$, $\delta>0$ and $\delta\,'>0$ are given. Consider natural positive numbers $d_\nu, \nu=1,2,...,N$, such that, $d_1 + d_2 +...+ d_N=D=$ dim $\mathcal{H}$, and moreover assume that, for all $\nu=1,2...,N$,
$$ d_\nu > D - \frac{\epsilon^2\, \delta\, \delta'\, D\, (D+1) }{N^3}.$$

Suppose $H:  \mathcal{H} \to \mathcal{H}$ is self-adjoint, {\bf the unitary vector $\psi_0\in \mathcal{H}$ is fixed}, and
$\psi_t = e^{-\,i\,t\,H} (\psi_0),$
$t \geq 0.$

Then, for $(1-\delta)$-most of the decompositions $\mathcal{D} \in \Delta(d_1,d_2,...,d_N, \mathcal{H} )$, the inequalities
$$ |\, \,|  E_{\psi_t} (P_{\mathcal{H}_\nu}) -\frac{d_\nu}{D}\,\,|  =  |\, \,| P_\nu (\mathcal{D})\, \psi_t\,|^2 -\frac{d_\nu}{D}\,\,|< \epsilon\, \sqrt{\frac{d_\nu}{N\, D}}   \,\,\,\,\,\,\,\,\,\,\,\,( \nu=1,2,...,N)\,\,$$
are true for $(1- \delta\,')$-most of the large times.

The estimates depend on  the initial condition $\psi_0$.

\end{theorem}

{\bf Proof:} We denote
$$f_{\nu} (\mathcal{D})\,=\,\lim_{T \to \infty} \,\, \frac{1}{T}\, \int_0 ^T \, (\, | P_\nu (\mathcal{D})\, \psi_t\,|^2 -\frac{d_\nu}{D}   )^2 \, dt.$$

From Corollary \ref{a8}, for each $\nu$

$$ w_\Delta (\,\{\mathcal{D} \in \Delta\,:\, f_\nu (\mathcal{D} ) \geq \,\frac{\epsilon^2\, \delta\, '\, d_\nu}{D\, N^2} \}\,) \leq $$
$$ \frac{d_\nu\, (D- d_\nu)}{D^2 \, ( D+1 )}\,\frac{D\, N^2}{\epsilon^2 \delta\,'\,d_\nu} =\frac{N^2\, (D- d_\nu)}{D \, ( D+1 )\, \epsilon^2\, \delta\, '}   . $$

Therefore, there exists a set $S\subset \Delta$ such that
$$ w_\Delta (S)\geq 1- \frac{N^3\, (D- d_\nu)}{D \, ( D+1 )\, \epsilon^2\, \delta\, '}> 1 - \delta,$$
and, at the same time $ f_\nu (\mathcal{D} )< \frac{\epsilon^2\, \delta\, '\, d_\nu}{D\, N^2}, $ for all $\mathcal{D}\in S$
and all $\nu=1,2...,N.$

Now, taking in Lemma \ref{a5} $\rho=\frac{\epsilon^2\, \delta\, '\, d_\nu}{D\, N^2},$ and $\gamma=\frac{\epsilon^2\, d_\nu}{D\, N},$ we get  for all $\mathcal{D}\in S$
and all $\nu=1,2,...,N$
$$  |\, \,| P_\nu (\mathcal{D})\, \psi_t\,|^2 -\frac{d_\nu}{D}\,\,|< \epsilon\, \sqrt{\frac{d_\nu}{N\, D}}   \,\,\,\,\,\,\,\,\,\,\,\,( \nu=1,2,...,N),$$
for $(1- \frac{\delta \,' }{N})$ most of the large times.

Therefore, the above inequalities for all $\nu=1,2,..,N$, are true for $(1- \delta \,' )$ most of the large times.

\qed

\medskip

Note that the mean value $f_{\nu} (\mathcal{D})\,$ depends of the Hamiltonian $H$ but the bounds of last theorem does not depend on $H$.

\bigskip

\section{Uniform estimates} \label{uni}

\medskip

In this section we will refine the last result considering uniform estimates which are independent of the initial condition
$\psi_0$ (for the time evolution associated to the fixed Hamiltonian $H: \mathcal{H} \to \mathcal{H}$).

\medskip

Suppose $\epsilon>0$, $\delta>0$ and $\delta\,'>0$ are given. Consider natural positive numbers $d_\nu, \nu=1,2,...,N$, such that, $d_1 + d_2 +...+ d_N=D=$ dim $\mathcal{H}$

We denote for each $\psi_0 \in \mathcal{H}$, where $|\psi_0|=1$, and   $  \mathcal{D}\in \Delta=\Delta(d_1,...,d_N; \mathcal{H})$
$$f_{\nu} (\psi_0,\mathcal{D})\,=\,\lim_{T \to \infty} \,\, \frac{1}{T}\, \int_0 ^T \, (\, | P_\nu (\mathcal{D})\, \psi_t\,|^2 -\frac{d_\nu}{D}   )^2 \, dt,$$
where $\psi_t= e^{-i\,t\, H} (\psi_0)$ (see Lemma \ref{a6}).

\begin{lemma} \label{a9} Suppose are given $\epsilon>0$ and $\delta'>0.$  Assume there exists non-negative   continuous functions $g_\nu: \Delta \to \mathbb{R}$, $\nu=1,2,..,N$, and $K>0$,
such  that

\begin{equation} \label{pp1} a) f_\nu( \psi_0 , \mathcal{D})\leq g_\nu,\, \,\text{for all}\,\,\,  \mathcal{D} \in \Delta\, \,\text{ and for all}\, \,\,\psi_0 \in \mathcal{H} \text{ with}\,\,|\psi_0|=1,
\end{equation}

\begin{equation} \label{pp2}
b) \int_\Delta g_\nu (\mathcal{D})\, d w_\Delta (\mathcal D)\, <K .
\end{equation}

Suppose $\delta$ is such that
\begin{equation} \label{d1}
1\,>\, \delta\geq \frac{K\,D\, N^3}{\epsilon^2\, \delta '\, d_\nu} , \nu=1,2,..,N.
\end{equation}

Then, for $(1-\delta)$ most of the $\mathcal{D} \in \Delta$ we have
\begin{equation} \label{d2}
|\,\,| P_\nu (\mathcal{D})\, \psi_t\,|^2 -\frac{d_\nu}{D} \,| \leq \epsilon\, \sqrt{\frac{d_\nu}{N\, D}},\,\,\nu=1,2,..,N ,
\end{equation}

for $(1- \delta ')$-most of the large times and for any $\psi_0 \in \mathcal{H}$ with $|\psi_0|=1$.

\end{lemma}

{\bf Proof:} Note that
$$w_\Delta (\,\{ \mathcal{D}\in \Delta\,:\, g_\nu (\Delta)\geq \delta ' \, \epsilon^2\, \frac{d_\nu}{N^2 \, D} \, \}\,< K\, \frac{N^2 \, D}{\delta '\, \epsilon^2 d_\nu} \,<\,  \frac{\delta}{N}, \,\nu=1,2,..,N .$$

Therefore, there exists a subset $E \subset \Delta$ such that $w_\Delta (E) < 1 - \delta$  and $g_\nu (\Delta)< \delta ' \, \epsilon^2\, \frac{d_\nu}{N^2 \, D}$, for all $\Delta \in E$ and all $\nu=1,2...,N.$

The conclusion is: if $\Delta\in E$, then,  $ f_\nu (\psi_0,\mathcal{D})\, < \delta ' \, \epsilon^2\, \frac{d_\nu}{N^2 \, D}$, for all  $\nu=1,2...,N,$
and all $\psi_0 $ with norm $1$.

The proof of the claim now follows from the reasoning of Theorem \ref{t1} and Lemma \ref{a5}.

\qed

Note that in order to have $\delta$ in expression (\ref{d1})   small it is necessary that all $d_\nu$ are large.
\medskip

We assume now several hypothesis on $H$
Consider a certain orthogonal basis of eigenvectors $\phi_1,\phi_2,...,\phi_D$ of $H$. We denote by $E_j$, $j=1,2,..,D$ the corresponding eigenvalues.
\medskip

We assume hypothesis $\mathfrak{N\,\,R}$ which says

a) $H$ is not degenerate, that is, $E_\alpha\neq E_\beta$, for $\alpha\neq \beta$,

and

b) $H$ has no resonances, that is, $E_\alpha-E_\beta \neq E_{\alpha '} - E_{\beta '}$, unless $\alpha= \alpha'$ and $\beta=\beta'$, or, $\alpha= \beta$ and $\alpha'=\beta'$.

\begin{lemma} \label{a10}

$$f_{\nu} (\psi_0,\mathcal{D})\leq \max_{1\leq \alpha \neq \beta \leq D}  | < \phi_\alpha ,P_\nu ( \mathcal{D})  \phi_\beta>|^2 +  \max_{1\leq \alpha \leq D}  ( < \phi_\alpha , P_\nu ( \mathcal{D}) \, \phi_\alpha\,>- \frac{d_\nu}{D})^2,$$
for all $\psi_0\in \mathcal{H}$, such that $|\psi_0|=1$, and for all $\mathcal{D} \in \Delta(d_1,...,d_N; \mathcal{H})$ and all $\nu=1,2...,N.$

\end{lemma}

{\bf Proof:} Suppose $\psi_0= \sum_{\alpha =1}^D \,c_\alpha\, \phi_\alpha$. Then,
$$\psi_t= \sum_{\alpha =1}^D \,c_\alpha\, e^{ -i\, t\,E_\alpha }\phi_\alpha , t \geq 0,$$

and,

$$|\,P_\nu ( \mathcal{D})  \psi_t|^2=< \psi_t ,P_\nu ( \mathcal{D})  \psi_t>  =$$
$$\sum_{1\leq \alpha, \beta \leq D} c_\alpha\,\overline{c}_ \beta e^{ -i t\,(E_\alpha- E_\beta )}\,\,\, \,\,\,\,\,< \phi_\alpha ,P_\nu ( \mathcal{D})  \phi_\beta>.$$

Therefore,
$$(\,|\,P_\nu ( \mathcal{D})  \psi_t|^2\,-\frac{d_\nu}{D}\,)^2 = $$
$$\sum_{1\leq \alpha, \beta, \gamma,\delta\leq D }^D\, c_\alpha\,\overline{c}_\beta \,c_\gamma\overline{c}_\delta\, e^{ -i t\,[\,(E_\alpha- E_\beta )\,- (E_\delta- E_\gamma)\,]}\,< \phi_\alpha ,P_\nu ( \mathcal{D})  \phi_\beta>\,< \phi_\gamma ,P_\nu ( \mathcal{D})  \phi_\delta>-$$
$$2\,\frac{d_\nu}{D}\,\sum_{1\leq \alpha, \beta \leq D} c_\alpha\,\overline{c}_ \beta e^{ -i t\,(E_\alpha- E_\beta )}\,\,< \phi_\alpha ,P_\nu ( \mathcal{D})  \phi_\beta>\,+\, \frac{d_\nu^2}{D^2}.$$

Using the above expression in the computation of integral $f_{\nu} (\psi_0,\mathcal{D})$ will remain just the terms where the coefficient of $t$ is zero.
By hypothesis, this will happen just when $\alpha= \delta$ and
$\beta=\gamma$, or, $\alpha=\beta$ and $\gamma=\delta$.

Note that the case $\alpha=\beta=\gamma=\delta$ is counted twice in the estimation.

Therefore,
$$f_{\nu} (\psi_0,\mathcal{D})  =\sum_{1\leq \alpha, \beta\leq D}\, |c_\alpha|^2\,|c_\beta|^2 \,\,\,\,|\,< \phi_\alpha ,P_\nu ( \mathcal{D})  \phi_\beta>\,|^2\,+$$
$$
\sum_{1\leq \alpha, \gamma\leq D}\, |c_\alpha|^2\,|c_\gamma|^2 \,\,< \phi_\alpha ,P_\nu ( \mathcal{D})  \phi_\alpha>\,< \phi_\gamma ,P_\nu ( \mathcal{D})  \phi_\gamma> -$$
$$\sum_{1\leq \alpha\leq D}\, |c_\alpha|^4\,|\,< \phi_\alpha ,P_\nu ( \mathcal{D})  \phi_\alpha>\,|^2\,- 2 \, \frac{d_\nu}{D}\,\sum_{1\leq \alpha\leq D}\, |c_\alpha|^2\,< \phi_\alpha ,P_\nu ( \mathcal{D})  \phi_\alpha>\,\,+\frac{d_\nu^2}{D^2},
$$
because
$\,< \phi_\gamma ,P_\nu ( \mathcal{D})  \phi_\delta>\,= \overline{< \phi_\delta ,P_\nu ( \mathcal{D})  \phi_\gamma>} $.

Finally, putting together the first and third terms
$$f_{\nu} (\psi_0,\mathcal{D})  =\sum_{1\leq \alpha \neq \beta\leq D}\, |c_\alpha|^2\,|c_\beta|^2 \,\,\,\,|\,< \phi_\alpha ,P_\nu ( \mathcal{D})  \phi_\beta>\,|^2\,+$$
$$(\,\sum_{1\leq \alpha\leq D}\, |c_\alpha|^2\,\,< \phi_\alpha ,P_\nu ( \mathcal{D})  \phi_\alpha>\,\,- \, \frac{d_\nu}{D}\,)^2\,.$$

By the other hand,
$$ \sum_{1\leq \alpha \neq \beta\leq D}\, |c_\alpha|^2\,|c_\beta|^2 \,\,\,\,|\,< \phi_\alpha ,P_\nu ( \mathcal{D})  \phi_\beta>\,|^2\,\leq $$
$$\max_{1\leq \alpha \neq \beta\leq D}\,|\,< \phi_\alpha ,P_\nu ( \mathcal{D})  \phi_\beta>\,|^2\, \sum_{1\leq \alpha ,\beta\leq D}\, |c_\alpha|^2\,|c_\beta|^2=$$
$$\max_{1\leq \alpha \neq \beta\leq D}\,|\,< \phi_\alpha ,P_\nu ( \mathcal{D})  \phi_\beta>\,|^2\, (\,\sum_{1\leq \alpha \leq D}\, \,|c_\alpha|^2\,)^2=$$
$$\max_{1\leq \alpha \neq \beta\leq D}\,|\,< \phi_\alpha ,P_\nu ( \mathcal{D})  \phi_\beta>\,|^2,$$
because $|\psi_0|=1$.

By the same reason
$$\,|\,\sum_{1\leq \alpha\leq D}\, |c_\alpha|^2\,\,< \phi_\alpha ,P_\nu ( \mathcal{D})  \phi_\alpha>\,\,- \, \frac{d_\nu}{D}\,|\,=$$
$$|\,\sum_{1\leq \alpha\leq D}\, |c_\alpha|^2\,(\,< \phi_\alpha ,P_\nu ( \mathcal{D})  \phi_\alpha>\,\,- \, \frac{d_\nu}{D})\,|\leq$$
$$\max_{1\leq \alpha\leq D}\,|\,\,< \phi_\alpha ,P_\nu ( \mathcal{D})  \phi_\alpha>\,\,- \, \frac{d_\nu}{D}\,|.$$

\qed

\medskip

Now, we define for each $\nu=1,2,...,N$  the continuous function $g_\nu (\mathcal{D}): \Delta(d_1,...,d_N; \mathcal{H}) = \Delta \to \mathbb{R} $ given by
 \begin{equation} \label{gnu} g_\nu (\mathcal{D}) =  \max_{1\leq \alpha \neq \beta\leq D}\,|\,< \phi_\alpha ,P_\nu ( \mathcal{D})  \phi_\beta>\,|^2\,+\max_{1\leq \alpha\leq D}\,|\,\,< \phi_\alpha ,P_\nu ( \mathcal{D})  \phi_\alpha>\,\,- \, \frac{d_\nu}{D}\,|^2.\end{equation}

We point out that for each $\mathcal{D}$ the expression $g_\nu(\mathcal{D})$ depends just on $H$ because as the $E_\alpha$ are all different the eigenvector basis is unique up to a changing in order and multiplication by scalar of modulus one.

Now we need a fundamental technical Lemma.

\begin{lemma} \label{88}

There exist a constant $C_1>0$ such that
$$ \int_\Delta g_\nu (\mathcal{D}) w_\Delta (\mathcal{D}) < \frac{10 \log D}{D}, \,\,\nu=1,2,...,N,$$

if, $C_1 \log D \,< \, d_\nu\,<\, \frac{D}{C_1}.$

\end{lemma}

Note that if $D$ is large there is a lot of room for the values $d_\nu$ to be able to satisfy last inequality. We will prove this fundamental  lemma in the next sections.

\medskip

If we assume the Lemma is true, then:

\begin{theorem} \label{main}
Given $ \epsilon, \delta>0$ and $\delta  '>0$, take $d_1,d_2,...,d_N$ such that, if $D=d_1+...+d_N$, $N>0$, then the following inequalities are true
$$ \max \, (\,C_1 , \frac{10 N^3}{\epsilon\, \delta \,\delta'}\,)\, \log D \,< d_\nu\,<\, \frac{D}{ C_1}, \,\,\nu=1,2,..,N, $$
where $C_1$ comes from Lemma \ref{88}.

Assume that $\mathcal{H}$ is a Hilbert space of dimension $D$ and $H: \mathcal{H} \to \mathcal{H}$ is a self-adjoint Hamiltonian without resonances and degeneracies, then, for $(1-\delta)$ most of the decompositions $\mathcal{D} \in \Delta(d_1,...,d_N;\mathcal{H})$ the system of inequalities
$$|\,\,\,|\,P_\nu ( \mathcal{D})  \psi_t\,|^2\,-\frac{d_\nu}{D}\,|\,  < \, \epsilon\, \sqrt{\frac{d_\nu}{N\, D}},\,\, \nu=1,2,...,N$$
are  true for most of the $(1-\delta ')$ large times and for any initial condition $\psi_0\in \mathcal{H}$, $|\psi_0|=1$.

\end{theorem}

{\bf  Proof:} By hypothesis and Lemma \ref{88} we get

$$ \int_\Delta g_\nu (\mathcal{D}) w_\Delta (\mathcal{D}) < \frac{10 \log D}{D}, \,\,\nu=1,2,...,N.$$

The claim follows from Lemma \ref{a9} by taking $K =  \frac{10 \log D}{D}.$
\qed

\bigskip

{\bf Main conclusion:}

As we said before, for a given fixed subspace   $\mathcal{H}_\nu$  of $\mathcal{H}$, the observable $P_{\mathcal{H}_\nu  }$ (the orthogonal projection on $\mathcal{H}_\nu$)
is such that the mean value $E_{\psi_t} (P_{\mathcal{H}_\nu})$  of the state $\psi_t$ is $<P_{\mathcal{H}_\nu  }(\psi_t), \psi_t>= |P_{\mathcal{H}_\nu  }(\psi_t)\,|^2.$

For a fixed Hamiltonian $H$ acting on a Hilbert space $\mathcal{H}$ of dimension $D$
the main theorem gives lower bound conditions on the dimensions $d_\nu$, $\nu=1,2,..,N$, of the different
$\mathcal{H}_\nu $ of a $(1-\delta)$-generic orthogonal decomposition $\mathcal{D}$ of the form $\mathcal{H}\,=\, \mathcal{H}_1\, \oplus ...\oplus \mathcal{H}_N$, in such way that the dynamic time evolution $\psi_t$, obtained from any fixed initial condition $\psi_0$,
for most of the large times $t$, has the property that the projected component $P_\nu (\mathcal{D})\, (\psi_t)\,=\,P_{\mathcal{H}_\nu  }(\psi_t)$
 is almost uniformly distributed (in terms of expected value)  with respect to the relative dimension size $\frac{d_\nu}{D}$ of $\mathcal{H}_\nu.$ In this way there is an approximately  uniform spreading of $\psi_t$ among the different $\mathcal{H}_\nu$ of the decomposition $\mathcal{D}$.
\bigskip

\section{Proof of Lemma \ref{88}}

The Lemmas \ref{22} and \ref{Lem23} will permit to reduce the integration problem from the unitary group to a problem in the real line.

\medskip

We will need first an auxiliary lemma. We denote by $S^k$ the unitary sphere in $\mathbb{R}^{k+1}$ and $S^k_r$ the sphere of radius $r>0$ in $\mathbb{R}^{k+1}.$ We consider the usual metric on them.

The next lemma is a classical result on Integral Geometry (see \cite{San}). We will provide a simple proof in  Appendix 2.

\begin{lemma} \label{20} Suppose $X$ is a Riemannian compact manifold, $f:X \to \mathbb{R}$ a $C^\infty$-function and $g: \mathbb{R} \to \mathbb{R}$ a continuous function. We define
$$G(v)=\, \int_{f\leq v} (g \circ f)\,  \lambda,$$
where $\lambda$ is the volume form on $X$. Suppose that $a\in \mathbb{R}$ is a regular value of $f$. Then, $G$ is differentiable at $v=a$ and
$$ \frac{d G}{dv} (a)\,=\,g(a)\, \int_{X_a} \frac{ \lambda_a}{|\, \,\text{ grad}\, f\,|},$$
where $X_a$ is the level manifold  $f=a$ and $\lambda_a$ is the induced volume form in $X_a$.

\end{lemma}

\medskip

\begin{corollary} \label{Cor21}
Given positive integers $d,D$, where $1<d<D-1$, denote by $S$ the unitary sphere on $\mathbb{R}^{2\, D},$ with the usual metric. Define
$$f(x)= x_1^2 +...+x^2_{2\,d }, \,\text{where}\,x \in S  \,\,\text{and}\,\,\,g:\mathbb{R} \to \mathbb{R}\,\,\,\text{is a continuous function}.$$

Suppose
$$G(v)=\, \int_{f\leq v} (g \circ f)\, d \lambda,$$
then
$G$ is of class $C^1$ and
$$ \frac{d G}{dv} (v)\,=\,\frac{2\, \pi^D}{(d-1)\,!\,\,(D -d-1)\,!}\, g(v)\, v^{d-1} \, (1-v)^{D-d-1},\,\,\,\text{if}\,\,0\leq v\leq 1,$$
and $ \frac{d G}{dv} (v)\,=0$, if $v<0$ or $v>1$.

\end{corollary}

{\bf Proof:} For  $x_1^2 +...+x^2_{2\,d }=v$ we have
$$ \text{grad}\,f(x)\,=\, 2\,( \,(1-v) x_1,...,(1-v) x_{2d}, -v\, x_{2d+1},...,-v\, x_{2\,D}\,).$$

Then, $|\text{grad}\,f(x)|=2\, \sqrt{v\, (v-1)}$, which is constant over $S_v=\{f=v\}$. Note that
$$S_v = S_{\sqrt{v}}^{2 d-1}\,\times S_{\sqrt{1-v}}^{2\,(D -d)\,-1},\,\,\,0<v<1.$$

From last Lemma and from the above expression it follows that (remember that vol $(S_r^{2n-1})\,= \,\frac{\,2 \, \pi^n}{(n-1)\, !}
\,r^{2n-1}$)
$$  \frac{d G}{dv} (v)\,=\,g(v)\,\frac{1}{\,2\, \sqrt{v\,(1-v)}\,}\, \,\frac{2\,\pi^d\,(\sqrt{v})^{2\,d-1}}{\,(d-1)\,!}\,\,  \,\frac{2\, \pi^{D-d} (\sqrt{(1-v)})^{2\,(D-d)\,-1}}{\,(D-d-1)\,!\, }           =\,$$
$$ \frac{2\,\pi^D\,v^{d-1}\,(1-v)^{D-d-1}\,}{\,(d-1)\,!\,\,\,\,\, (D-d-1)\,!},\,\,\,\,\,\,\,\,0<v<1.$$

In the case $v<0$ or $v>1$, we have that $G$ is constant. Finally, as $S_0$ and $S_1$ are submanifolds of $S$ we have that $G$ is continuous for $v=0$ and $v=1$.

\qed

From now on we fix $\nu$, where $1\leq \nu\leq N$, and we define
$$e_{\alpha,\beta}( \mathcal{D})= <\, \phi_\alpha, P_\nu (\mathcal{D})\, \phi_\beta\,>,  \,\,\,\mathcal{D} \in \Delta, \,\,1\leq \alpha,\,\beta\,\leq D,\,\,\,e_{\alpha,\beta}:\Delta\to \mathbb{C},$$
where $\phi_1,...,\phi_D$ is the orthonormal basis for $\mathcal{H}$ which were fixed in section \ref{uni}.

\medskip

\begin{lemma} \label{22} Suppose $1< d_\nu<D-1$. Let $a \geq 0$ be such $\sqrt{a} < \frac{d_\nu}{D}$ and
$\sqrt{a}+  \frac{d_\nu}{D}<1.$ Then, the probability such that
$(e_{\alpha,\beta}\,-\frac{d_\nu}{D} )^2\, \geq \alpha$ is
$$ \frac{(D-1)\,!\,}{\,(d_\nu-1)\,!\, (D-d_\nu-1)\,!}\,\int_{[0,\,\frac{d_\nu}{D}-\sqrt{a}]\cup [\frac{d_\nu}{D}+\sqrt{a},\,1]} u^{d _\nu -1}\,(1-u)^{D-d_\nu-1}\,d u.$$
\end{lemma}
\medskip

\begin{lemma} \label{Lem23} Suppose $1< d_\nu<D-1$. Let $\alpha\neq \beta$ and $0\leq a\leq 1/4$. Then, the probability such that $|\,e_{\alpha,\beta}\,|^2\, \geq a$ is
$$ \frac{(D-1)\,!\,}{\,(d_\nu-1)\,!\, (D-d_\nu-1)\,!}\,\int_{1/2\,-\, \sqrt{1/4-a}}^{1/2\,+\, \sqrt{1/4-a}}\,\frac{(w\,(1-w)-a)^{D-2}}{w^{D-d_\nu -1}\,(1-w)^{d_\nu-1}} d w.$$

\end{lemma}
\medskip

{\bf Proof of Lemma \ref{22}:} We just have to consider the case $\nu=1$. We write $d=d_1$ and denote by $P$ the orthogonal projection of $\mathcal{H}$ over $\mathbb{C} \phi_1+...+ \mathbb{C} \phi_d.$

We denote by $p:\mathbb{U} \to \Delta$ the projection defined in the beginning of section \ref{sec2}, where
$\mathbb{U}$ denotes the group of unitary transformations of $\mathcal{H}.$

If $U\in \mathbb{U}$, then

$ e_{\alpha,\alpha}(p(U))=<\phi_\alpha,$ orthogonal projection of $\phi_\alpha$ in $\mathbb{C} U(\phi_1)+...+ \mathbb{C}\,U( \phi_d)>=$

$\,\,\,\,\,\,\,\,\,\,\,\,\,\,\,\,\,\,\,\,\,\,\,\,\,\,\,\,\,\,\,\,\,\,\,\,\,\,\,\,\,\,\,\,\,\,\,\,\,\,\,\,<U^{-1} (\phi_\alpha), P (U^{-1} \phi_\alpha)>.$

Denote $q:\mathbb{U} \to S,$\,\,  where $q(U)= U(\phi_\alpha),\,\,\,U \in \mathbb{U}$ and
$\sigma:S \to \mathbb{R},$\,\,where  $\sigma(\phi)= <\phi, P(\phi)>,$\,\,\,$\phi \in S,$ and where $S$ is the unitary sphere of $\mathcal{H}.$
\medskip

Then, we get the following commutative diagram:
$$  \,\,\,\,\,\text{inverse} \,\,\,\,\,\,\,\,\,$$
$$ \mathbb{U} \,\,\,\,\,\to \,\,\,\,\,\,\,\,\,\mathbb{U}$$
$$ p \downarrow \,\,\,\,\,\,\, \,\,\,\,\,\, \,\,\,\,\, \,\,\,\,\downarrow q$$
$$ \Delta \,\,\,\,\,\,\, \,\,\,\,\,\, \,\,\,\,\, \,\,\,\,\,\,\,\,\,S$$
$$ e_{\alpha,\alpha} \searrow \,\,\,\,\,\,\,\, \,\,\swarrow \sigma$$
$$   \,\,\,\,\,\, \,\,\mathbb{R}$$

As the inverse preserves the metric, it follows from Lemma \ref{Lem1} a) that the probability of
$ e_{\alpha,\alpha} \leq b$ is equal to the probability that $\sigma\leq b$. Note that the metric on $S$ as quotient of $\mathbb{U}$ is the same as the induced by $\mathcal{H}$ because $\mathbb{U}$ acts transitively on $S$.

It will be more easy to make the computations via the right hand side of the diagram.

We identify $\mathcal{H}$ with $\mathbb{C}^D= \mathbb{R}^{2\, D}$, via $\phi_1,\phi_2,...,\phi_D$. Then $S$ is identified with the unitary sphere in
$\mathbb{R}^{2\, D}$, also denoted by $S$, and
$$ \sigma:S \to \mathbb{R},\,\, \,\,\sigma(x) = x_1^2+...+ x_{2 \,d}^2 , \,\,\, x \in S.$$

Therefore, by Corollary \ref{Cor21} with $g=1$ we get
$$ \frac{d\, (\text{Vol}\, (\sigma\leq v)\,)}{d\,v} \,=\,\frac{2\, \pi^D}{(d-1)\,!\,\,(D -d-1)\,!}\, v^{d-1} \, (1-v)^{D-d-1},\,\,\,\text{if}\,\,0\leq v\leq 1,$$
and
$$\frac{d\, (\text{Vol}\, (\sigma\leq v)\,)}{d\,v} \,=0,$$
if $v<0$ or $v>1$.

Now, we normalize dividing by vol $S=  \frac{\, 2\,\pi^D}{(D-1)\,!}$ and we get
$$ \frac{d\, (\text{prob}\, (\sigma\leq v)\,)}{d\,v} \,=\,\frac{(D-1)\,!}{(d-1)\,!\,\,(D -d-1)\,!}\, v^{d-1} \, (1-v)^{D-d-1},\,\,\,\text{if}\,\,0\leq v\leq 1.$$

As $(e_{\alpha,\alpha}\,-\frac{d}{D} )^2\, \geq a$ is equivalent to
$$ e_{\alpha,\alpha}\,\geq \frac{d}{D} \, +  \sqrt{a},\,\,\,\text{or}\,\,\, e_{\alpha,\alpha}\,\leq \frac{d}{D} \, -  \sqrt{a} ,$$
we get that the probability of $(e_{\alpha,\alpha}\,-\frac{d}{D} )^2\, \geq a$ is equal to
the probability of $\sigma \geq \frac{d}{D} \, +  \sqrt{a}$ or $\sigma \leq \frac{d}{D} \, -  \sqrt{a}$. From this follows that
the probability of $(e_{\alpha,\alpha}\,-\frac{d}{D} )^2\, \geq a$ is equal to
$$\frac{(D-1)\,!}{(d-1)\,!\,\,(D -d-1)\,!}\,\,[\,\int_{\frac{d}{D} +\sqrt{a} }^1v^{d-1} \, (1-v)^{D-d-1}dv+ \int^{\frac{d}{D} -\sqrt{a} }_0 v^{d-1} \, (1-v)^{D-d-1}dv\,].$$

Observe that $\sigma=$ constant is an analytic subset of $S$ and therefore the associated probability is zero. The case $a=0$ is trivial.
\qed

\medskip

{\bf Proof of Lemma \ref{Lem23}:} We just have to consider the case $\nu=1$. Take $d=d_1$ and as before we denote by $P$ the orthogonal projection of $\mathcal{H}$ over $\mathbb{C} \phi_1+...+ \mathbb{C} \phi_d.$
Once more we denote by $p:\mathbb{U} \to \Delta$ the projection defined in the beginning of section \ref{sec2}.

If $U\in \mathbb{U}$, then

$ e_{\alpha,\beta}(p(U))=<\phi_\alpha,$ orthogonal projection of $\phi_\beta$ in $\mathbb{C} U(\phi_1)+...+ \mathbb{C}\,U( \phi_d)>=$

$\,\,\,\,\,\,\,\,\,\,\,\,\,\,\,\,\,\,\,\,\,\,\,\,\,\,\,\,\,\,\,\,\,\,\,\,\,\,\,\,\,\,\,\,\,\,\,\,\,\,\,\,<U^{-1} (\phi_\alpha), P (U^{-1} \phi_\beta)>.$

Denote $q_{\alpha,\beta} :\mathbb{U} \to S\times S,$\,\,  where $q_{\alpha,\beta}(U)= (\,U(\phi_\alpha), U(\phi_\beta)\,),\,\,\,U \in \mathbb{U}$, and  $S$ is the unitary sphere of $\mathcal{H}.$
\medskip

Denote by $M= q_{\alpha,\beta}(\mathbb{U})= \{(\phi,\psi)\in S \times S\,| \,\phi$ is orthogonal to $ \psi\,\}$.

Let $H_{\alpha,\beta} \subset \mathbb{U}$ the closed subgroup of the $U$ such that $U(\phi_\alpha)= \phi_\alpha$ and $U(\phi_\beta)= \phi_\beta$.

Then, $M = \mathbb{U}/H_{\alpha,\beta}$ and $q_{\alpha,\beta}:  \mathbb{U} \to M$ is the canonical projection.

The quotient metric on $M$ is the induced by $S\times S$ because $\mathbb{U}$ acts transitively on $M$.

\medskip

\newpage

Let $f:M \to \mathbb{C}$ given by $f(\phi, \psi) = <\,\phi, P( \psi)\,>.$
Then, we get the following commutative diagram:
$$  \,\,\,\,\,\text{inverse} \,\,\,\,\,\,\,\,\,$$
$$ \mathbb{U} \,\,\,\,\,\to \,\,\,\,\,\,\,\,\,\mathbb{U}$$
$$ p \downarrow \,\,\,\,\,\,\, \,\,\,\,\,\, \,\,\,\,\, \,\,\,\,\downarrow q_{\alpha, \beta}$$
$$ \Delta \,\,\,\,\,\,\, \,\,\,\,\,\, \,\,\,\,\, \,\,\,\,\,\,\,\,\,M$$
$$ e_{\alpha,\beta} \searrow \,\,\,\,\,\,\,\, \,\,\swarrow f$$
$$   \,\,\,\,\,\, \,\,\mathbb{C}$$

As the inverse preserves the metric of $\mathbb{U}$, it follows that the probability of
$ |e_{\alpha,\alpha}|^2 \leq a$ is equal to the probability that $|f|^2\leq a$ by Lemma \ref{Lem1} a).

Now consider $\varphi:M \to S$, such that $\varphi(\phi,\psi)=\psi$. This defines a $C^\infty$  locally trivial fiber bundle with fiber $S^{2 \, D-3}$. Indeed,
$E_\psi=\varphi^{-1} (\psi)$ is the unitary sphere of the subspace $\mathcal{H}_\psi$ which is is the orthogonal set to $\psi$ in $\mathcal{H}$.

Given $u\in \mathbb{R}$ denote:
$$ F_u(\psi)= E_\psi\,\cap\, \{|f|^2 \leq u  \},\,\,\,\psi \in S.$$

Then,

$$ \text{Vol}\, (\{|f|^2 \leq u  \})\,=\, \int_S \text{vol}_{E_\psi} (F_u (\psi))\,\,\, d S \,(\psi).$$

For each $\psi$ we get $\psi'\in \mathcal{H}$ via
$$ P(\psi)=\, c \psi + \psi', \,\,\text{where}\,\,c \in \mathbb{C}\,\,\, \text{and}\,\,\,\psi' \, \,\text{is orthogonal to}\,\,\psi.$$

Note that $\psi' \in \mathcal{H}_\psi$. Then
$$ f(\phi,\psi)=<\phi,P(\psi)>=<\phi,\psi'>,$$
and it follows that
$$F_u (\psi) = \{\phi\in E_\psi\,:\, \,| \,<\phi,\psi'>\,|^2\leq u\},\,\,u\in \mathbb{R},\,\,\psi\in S.$$

There exist an isomorphism identifying $\mathcal{H}_\psi= \mathbb{C}^{D-1}= \mathbb{R}^{2\, D -2}$ between Hilbert spaces which transform $\psi'$ in
$(|\psi'|,0,...,0)$. This isomorphism identifies $E_\psi$ with the unitary sphere $E$ on $\mathbb{R}^{2\, D-2}$ and $F_u (\psi)$ with the set

$$ \{ x \in E \,:\, |\psi'|^2 (\,x_1^2 + x_2^2\,) \leq u\}.$$

Now applying Corollary \ref{Cor21} with $D-1$ instead of $D$, $d=1$, $g=1$ and $v= \frac{u}{|\psi'|^2}$ we get
$$ \frac{d\,\text{Vol}_{E_\psi}  F_u(\psi)}{d\, u}  = \frac{2\, \pi^{D-1}}{(D-3)\,!} (1- \frac{u}{|\psi \,'|^2})^{D-3} \frac{1}{|\psi \,'|^2} =\frac{2\, \pi^{D-1}}{(D-3)\,!}  \frac{(|\,\psi\,'|^2- u)^{D-3}}{|\psi \,'|^{2\,(D-2)}} , $$
for all $\psi \in S$ and $0<u\leq |\psi\,'|^2$, and
$$ \frac{d\,\text{Vol}_{E_\psi}  F_u(\psi)}{d\, u}  = 0$$
if $|\psi\, '|^2 \leq u\leq 1$, for any $\psi\in S$.

Then we get that $ \frac{d\,\text{Vol}_{E_\psi}  F_u(\psi)}{d\, u}$ is a continuous function of $(u,\psi)$ for $0<u\leq 1$ and $\psi\in S$. As $S$ is compact
we can take derivative inside the integral and we get

$$ \frac{d\,\text{Vol} (|f|^2 \leq u)}{d\, u}= \int_S \,  \frac{d\,\text{Vol}_{E_\psi}  F_u(\psi)}{d\, u} \, d S(\psi)$$
for any $0<u\leq 1.$

By the definition of $\psi\,'$ it is easy to see that $|\psi\,'|^2 = |P(\psi)|^2 \,(1- |P(\psi)|^2)$.

Now we consider $g_u: \mathbb{R} \to \mathbb{R} $ where
$$ g_u(w) = \frac{(w\, (1-w) - u)^{D-3}}{(w\, (1-w))^{D-2}} $$
if $u \leq w\, (1-w),$
and $g_u(w)=0$ in the other case.

$g_u(w)$ is a continuous function of $u$ and $w$ when $0<u\leq 1$, $0\leq w \leq 1$.

From this follows that
$$ \frac{d\,\text{Vol} (|f|^2 \leq u)}{d\, u}= \frac{2\, \pi^{D-1}}{(D-3)\,! } \int_S \,  (\,g_u \circ |P(\psi)|^2 \,) \, d S(\psi)$$
for any $0<u\leq 1.$

Now we normalize dividing by Vol $(M)= \frac{ 2\, \pi^{D-1 }}{(D-2)\, !}\,\frac{ 2\, \pi^{D }}{(D-1)\, !}$
and we get
\begin{equation} \label{Lu}
\frac{d\,\text{Prob} (|f|^2 \leq u)}{d\, u}= \frac{(D-1)\,!\, (D-2)}{(\, 2\,\pi^D)\, } \int_S \,  (\,g_u \circ |P(\psi)|^2 \,) \, d S(\psi)\end{equation}
for any $0<u\leq 1.$

Denote
$$ A(u,w) = \int_{|P(\psi)|^2 \leq w} (\,g_u \circ |P(\psi)|^2 \,) \, d S(\psi),$$
for any $0<u\leq 1$, $0\leq w\leq 1.$

By Corollary \ref{Cor21} we get
$$\int_S\,( \,g_u \circ |P(\psi)|^2 \,) \, d S(\psi)\, = A(u,1) = A(u,1) - A(u,0) = \int_0^1 \frac{\partial A}{\partial w}(u,w)\, dw,$$
for any $0<u\leq 1.$

Estimating $\frac{\partial A}{\partial w}$ by Corollary \ref{Cor21} and substituting in (\ref{Lu}) we finally get
$$
\frac{d\,\text{Prob} (|f|^2 \leq u)}{d\, u}= \frac{(D-1)\,!\, (D-2)}{(d-1)\,!\, (D-d-1)\, !\, } \int_{u \leq w\,(1-w)} \,  \frac{(w\, (1-w)- u)^{D-3} }{w^{D-d-1} \, (1-w)^{d-1}}  \, d w$$
for any $0<u\leq 1.$

If $u>1/4$, $w\,(1-w)<u$ for all $w$ and the integral is zero.

If $0<u\leq 1/4$, $u\leq w (1-w)$ is equivalent to
$$1/2 - \sqrt{1/4-u} \leq w\leq 1/2 + \sqrt{1/4-u}. $$

Then,
$$
\frac{d\,\text{Prob} (|f|^2 \leq u)}{d\, u}=
 $$
 $$\frac{(D-1)\,!\, (D-2)}{(d-1)\,!\, (D-d-1)\, !\, } \int_{1/2 - \sqrt{1/4-u} }^{1/2 + \sqrt{1/4-u}}  \,  \frac{(w\, (1-w)- u)^{D-3} }{w^{D-d-1} \, (1-w)^{d-1}}  \, d w,$$
if $0<u\leq 1/4$,
and
$$\frac{d\,\text{Prob} (|f|^2 \leq u)}{d\, u}=0
 $$
if $1/4\leq  u\leq 1.$

Finally, for $0< a\leq 1/4$
$$
\text{Prob} (|f|^2 \geq a)=
 $$
 $$\frac{(D-1)\,!\, (D-2)}{(d-1)\,!\, (D-d-1)\, !\, } \int_a^{1/4}\,du\,\int_{1/2 - \sqrt{1/4-u} }^{1/2 + \sqrt{1/4-u}}  \,  \frac{(w\, (1-w)- u)^{D-3} }{w^{D-d-1} \, (1-w)^{d-1}}  \, d w.$$

Considering the double integral in the region $a\leq u\leq w\, (1-w)$ we get
$$\text{Prob} (|f|^2 \geq a)=
 $$
 $$\frac{(D-1)\,!\, (D-2)}{(d-1)\,!\, (D-d-1)\, !\, }\int_{1/2 - \sqrt{1/4-u} }^{1/2 + \sqrt{1/4-a}}  \,dw\,\int_a^{w\,(1-w)}   \frac{(w\, (1-w)- u)^{D-3} }{w^{D-d-1} \, (1-w)^{d-1}}  \, d u=$$
$$\frac{(D-1)\,!\,}{(d-1)\,!\, (D-d-1)\, ! } \int_{1/2 - \sqrt{1/4-a} }^{1/2 + \sqrt{1/4-a}} \frac{(w\, (1-w)- a)^{D-2} }{w^{D-d-1} \, (1-w)^{d-1}}  \, dw.$$

The case $a=0$ is trivial.

\qed

\medskip

{\bf Remark:} Note that if $g:\Delta \to \mathbb{R}$ is a continuous function such that $0\leq g( \mathcal{D})\leq r$, for all $\mathcal{D} \in \Delta,$
then we get the estimate
$$ \int_\Delta g( \mathcal{D})\, w_\Delta  (\mathcal{D})\,=  \int_{g \geq a} g( \mathcal{D})\, w_\Delta  (\mathcal{D})\,+  \int_{g < a} g( \mathcal{D})\, w_\Delta  (\mathcal{D})\,\leq $$
$$ r\, \,\text{Prob}\,\,(g\geq a)+ \, a,$$
for $0\leq a\leq 1$.

Given positive integer numbers $d,D$ and $a \in \mathbb{R}$ such that
$$ 1<d<D-1,\,\,\,\,0\leq a \leq \frac{d^2}{D^2}\,\,\,\text{and}\,\,\,\frac{d}{D} + \sqrt{a} \leq 1$$
we define

$$I(d,D,a) = \frac{(D-1)\,!}{(d-1)\,!\, (D-d-1)\, ! } \int_{[0,\,\frac{d}{D} - \sqrt{a}] \cup [\frac{d}{D}+ \sqrt{a} , \,1] }\, u^{d-1}\,(1-u)^{D-d-1}\, du.$$

Below we will use the estimate $\theta=11/12.$

\begin{lemma} \label{Lem24} There exists a constant $C>4$, such that, if $a \geq 0,$ $d\geq 1$ $C\, \log D< d <\frac{D}{C}$ and $\frac{1}{D} < \sqrt{a} < \frac{d}{8\, D} $, then
$$ I(d,D,a) < \frac{D}{\sqrt{d}}\, e^{- \frac{\theta\,\, a\, \,D^2}{2\,d} }.$$

\end{lemma}

{\bf Proof:} Note that our hypothesis implies that $1<d<D-1$, $a^2 < \frac{d^2}{D^2}$ and $\frac{d}{D} + \sqrt{a} <1$.

\medskip

a) By Stirling formula, when $D\to \infty$, $d \to \infty$,  $D/d\to \infty$, we get that

$$ \frac{(D-1)\,!}{(d-1)\,!\, (D-d-1)\, ! }\sim \frac{1}{e} \, \sqrt{\frac{d}{2 \pi} } (\frac{d}{D})^{-d} (1- \frac{d}{D})^{d-D}.$$

As $\sqrt{\frac{1}{2 \pi}}\,<\,1$, there exists a constant $A$ such that if $D>A$, $d>A$ and $D/d>A$, we get

$$ \frac{(D-1)\,!}{(d-1)\,!\, (D-d-1)\, ! }\,< \, \frac{\sqrt{d}}{2 }  (\frac{d}{D})^{-d} (1- \frac{d}{D})^{d-D}.$$

If we take $C>A+1$ it follows from the hypothesis of the Lemma that $D>d\,C >d\,A$, $d>C \log D>C>A$ and $D-d>d\, C-d=d (C-1)> d\, A>A$.

b) The derivative of $u^{d-1}\, (1-u)^{D-d-1}  $ with respect to $u$ in $(0,1)$ is zero only on the point $u= \frac{d-1}{D-1}$  which is smaller than
$d/D$.

Moreover,
$$ \frac{d}{D}- \sqrt{a}<\frac{d}{D}- \frac{1}{D}=\frac{d-1}{D}< \frac{d-1}{D-1}.$$

Then,
$\frac{d-1}{D-1}\in ( \frac{d}{D}- \sqrt{a}, \frac{d}{D} )\subset  ( \frac{d}{D}- \sqrt{a}, \frac{d}{D} + \sqrt{a}).$

From this it follows that $u^{d-1}\, (1-u)^{D-d-1} $ takes its maximal values on the set $[0, \frac{d}{D}- \sqrt{a}]\,\cup\,
[\frac{d}{D}+ \sqrt{a},1]$ on the point $\frac{d}{D}- \sqrt{a}$ or on the point $\frac{d}{D}+ \sqrt{a}.$

Under our hypothesis, if $C>A+1$ we get that for $\epsilon=1$ or $-1$:

$$I(d,D,a) < \frac{\sqrt{d}}{2 } (\frac{d}{D})^{-d} (1- \frac{d}{D})^{d-D} (\,\frac{d}{D} + \epsilon \sqrt{a})^{d-1} (\,1-\frac{d}{D} - \epsilon \sqrt{a})^{D-d-1} =$$
$$ \frac{\sqrt{d}}{2 }\, \frac{(1 + \epsilon\frac{D}{d}  \sqrt{a})^{d} \,(1 - \epsilon\frac{D}{D-d}\sqrt{a})^{D-d}   }{ (\,\frac{d}{D} + \epsilon \sqrt{a})\,  (1-\,\frac{d}{D} - \epsilon \sqrt{a}) }.$$

\medskip

c) If $\epsilon=1$ with $C>4$, $C>A+1$ we get
$$ (\,\frac{d}{D} + \epsilon \sqrt{a})\,  (1-\,\frac{d}{D} - \epsilon \sqrt{a})= \frac{d}{D} + \sqrt{a}\, - \,\frac{d^2}{D^2} - 2\frac{d}{D} \sqrt{a} - a>$$
$$ \frac{d}{D} -  \, \frac{d^2}{D^2} - 2 \frac{d}{D} \sqrt{a} > \frac{d}{D} -  \, \frac{d^2}{D^2} - 2 \frac{d^2}{8\, D^2}  =\frac{d}{D} -  \, \frac{5\,d^2}{4\,D^2}> \frac{d}{D} (1 -\frac{5\,d}{4\,D}) >\frac{d}{2\,D}.$$
\medskip

If $\epsilon=1$ with $C>4$, $C>A+1$ one can show in the same way that
$$ (\,\frac{d}{D} + \epsilon \sqrt{a})\,  (1-\,\frac{d}{D} - \epsilon \sqrt{a})> \frac{d}{2\, D}.$$

\medskip

In this way we finally get that for $\epsilon=1$ or $\epsilon=-1$

$$I(d,D,a) <  \frac{\sqrt{d}}{2 } \frac{2 \, D}{d}  (\,1+ \epsilon \,\frac{D}{d} \sqrt{a})^{d} (\,1- \epsilon \frac{D}{D-d} \sqrt{a})^{D-d} =$$
$$  \frac{ D}{\sqrt{d}}  (\,1+ \epsilon \,\frac{D}{d} \sqrt{a})^{d} (\,1- \epsilon \frac{D}{D-d} \sqrt{a})^{D-d} .$$
\bigskip

Note that
$$  \frac{ D}{\sqrt{d}}  (\,1+ \epsilon \,\frac{D}{d} \sqrt{a})^{d} (\,1- \epsilon \frac{D}{D-d} \sqrt{a})^{D-d} =$$
$$  \frac{ D}{\sqrt{d}}\, \exp \,[ d\, \log (  1+ \epsilon \,\frac{D}{d} \sqrt{a})  + (D-d) \log (1- \epsilon \frac{D}{D-d} \sqrt{a} )]<$$
$$  \frac{ D}{\sqrt{d}}\,\exp\, [ d\, (\epsilon \,\frac{D}{d} \sqrt{a} - \frac{1}{2} \frac{D^2}{d^2} \, a +\frac{\epsilon}{3} \frac{D^3}{d^3} \, a^{3/2}  )  + (D-d) (- \epsilon \frac{D}{D-d} \sqrt{a} )\,].$$

This is so because $\log(1+x)=x- \frac{x^2}{2} + \frac{x^3}{3}+...$, for $|x|<1$, $\frac{D}{d} \sqrt{a}<1/8$,  and
$\frac{D}{D-d} \sqrt{a}<\frac{1}{24}.$

Therefore, if $C>4$ and $C>A+1$, then
$$I(d,D,a) <  \frac{ D}{\sqrt{d}} \exp [ \,-\frac{1}{2}  \,\frac{D^2}{d} \,a+ \frac{\epsilon}{3}  \,\frac{D^3}{d^2} \,a^{3/2} ],$$
for $\epsilon=1$ or $\epsilon=-1$.

Note that
$$ \frac{|\frac{\epsilon}{3}  \,\frac{D^3}{d^2} \,a^{3/2}|}{|-\frac{1}{2}  \,\frac{D^2}{d} \,a |} =
\frac{2}{3}  \,\frac{D}{d} \,a^{1/2}< \frac{2}{3}  \,\frac{D}{d} \,\frac{d}{8\, D} =\frac{1}{12}.$$

Therefore, if $C>4$ and $C>A+1$, we finally get
$$I(d,D,a) <  \frac{ D}{\sqrt{d}} e^{- \frac{\theta}{2}\, \frac{D^2}{d} \,a  }.$$

\qed

\medskip

Motivated by the Remark  before Lemma \ref{Lem24} we will choose a convenient choice of $a$.

\medskip

\begin{corollary} \label{Cor25}

There exist $C_0>4$ such that if $d$ and $D$ are such that
$ C_0 \log D < d < \frac{D}{C_0}$, then
$$ I(d,D,a)< \frac{1}{D^3 \sqrt{d}},$$
where
$a=\frac{8\, d\, \log D}{\theta \, D^2}.$

\end{corollary}

{\bf Proof:} Take $C_0>C$ (of Lemma \ref{Lem24}) and $C_0> 24^{2}$. Then,
$$ \sqrt{a} = \sqrt{\frac{8}{\theta}} \,  \frac{ \sqrt{d\, \log D}}{ D}< 3\,   \frac{ \sqrt{d\, \frac{d}{C_0}}}{ D}= \frac{3\,d}{ D\,\sqrt{C_0}}< \frac{3\,d}{ D\,24}= \frac{d}{ 8\, D\,},$$
because $\frac{8}{\theta}<9.$

Moreover,
$\sqrt{a}> \frac{ \sqrt{d\, \log D}}{ D}> \frac{1}{ D}.$

By Lemma \ref{Lem24}, we get that
$$ I(d,D,a) < \frac{D}{\sqrt{d}}\, e^{ -\, \frac{8\, \theta\, D^2}{2\,d} \,\frac{8}{\theta}\, \frac{d\, \log D}{D^2} }=\frac{D}{\sqrt{d}} \, e^{ -\,4\, \log D } = \frac{1}{ D^3 \sqrt{d}}.$$

\qed

\medskip

\begin{lemma} \label{Lem26}

Suppose $C_0$ is the constant of Corollary \ref{Cor25}. Given $1\leq \nu\leq N,$ suppose that $C_0\log D < d_\nu< \frac{D}{C_0},$ then,
$$ \int_\Delta \, \max_{1\leq \alpha\leq D} \,( <\,\phi_\alpha, P_\nu (\mathcal{D})\, \phi_\alpha\,>- \frac{d_\nu}{D}\,)^2 w_\Delta(\mathcal{D})\,<\, \frac{9\, d_\nu\, \log D}{D^2}.$$

\end{lemma}

{\bf Proof:} Suppose $a= \frac{8\, d_\nu\, \log D}{\theta \, D^2}$.

By Corollary \ref{Cor25} and Lemma \ref{22} (see also the beginning of the proof of Lemma \ref{Lem24}) we get that the probability of the above integrand to be great or equal to $a $
is smaller than
$ D\, \frac{1}{D^3\, \sqrt{d_\nu}} = \frac{1}{D^2 \sqrt{d_\nu} }$.

As we point out in the Remark before Lemma \ref{Lem24} the integral is smaller than
$$\frac{1}{D^2 \sqrt{d_\nu} }\,+\, \frac{8}{\theta}\, \frac{d_\nu \, \log D}{ D^2}.$$

Note that
$$ \frac{\frac{1}{D^2 \, \sqrt{d_\nu} }}{ \frac{d_\nu \, \log D}{ D^2}}\,=\, \frac{1}{d_\nu^{3/2} \,\log D}<9- \frac{8}{\theta}=\frac{3}{11},$$
because $d_\nu^{3/2} \,\log D> C_0^{3/2}\, (\log D)^{5/2} > C_0^{3/2}>8> \frac{11}{3}.$

Therefore,
$$\frac{1}{D^2 \sqrt{d_\nu} }\,+\, \frac{8}{\theta}\, \frac{d_\nu \, \log D}{ D^2}<  (9-\frac{8}{\theta} )\,\frac{d_\nu \, \log D}{ D^2}\,+\frac{8}{\theta} \, \frac{d_\nu \, \log D}{ D^2}\,= \frac{9\,d_\nu \, \log D}{ D^2} .$$

\qed

\medskip

In Lemma \ref{88} the function $g_\nu$ is defined as the sum of two terms (see expression (\ref{gnu}). The Lemma \ref{Lem26} takes care of the upper bound of the integral of the second term. Now we will estimate the upper bound for the first term (using the Remark done before Lemma \ref{Lem24}). First we need two lemmas.

\medskip

\begin{lemma} \label{Lem27} Suppose $\phi$ and $\psi$ are orthonormal and $E\subset \mathcal{H}$ is a subspace. Denote by $P$ the orthogonal projection of $\mathcal{H}$ over $E$.

Then, $ |\,<\phi\,,\,P(\psi)\,>\,|^2\leq 1/4$.

\end{lemma}

{\bf Proof:} If $\psi$ is orthogonal to $E$ or $\psi \in E$ we have that $<\phi,P(\psi)>=0.$

Suppose $\psi$ is not on $E$ and is also not orthogonal to $E$. Suppose $\psi=\psi_1+\psi_2$, where $\psi_1 $ is orthogonal to $E$ and $\psi_2 \in E$.

Let $\lambda=|\psi_1| $ and $\mu =|\psi_2|$, then $\psi_1=\lambda e_1$, $\psi_2= \mu \, e_2$, where $e_1$ and $e_2$ are orthonormal.

Denote by $\theta$ the orthogonal projection of $\phi$ over $\mathbb{C}\, e_1 + \mathbb{C}\, e_2$. Then,
$$ |\theta|\leq 1 \,\,\,\text{and}\,\, \,\,\alpha\,=\, <\phi, P(\psi)>= < \phi,\psi_2>= <\theta,\psi_2>.$$

Now, $<\phi\,,\,\psi>=0$ implies that
$$0 = <\phi,\psi_1>+<\phi,\psi_2>=<\theta,\psi_1>+<\theta,\psi_2>.$$

Suppose $\theta= a \, e_1 + b\, e_2$, then $|a|^2 + |b|^2\leq 1.$ By the other hand $1=|\psi|=|\psi_1 + \psi_2|= |\lambda|^2 + |\mu|^2$ and

$$ \alpha= <\theta, \psi_2>\,=\,b\, \overline{\mu} ,\,\,\, \,\, <\theta, \psi_1>\,=\,a\, \overline{\lambda},\,\,\, \,\,a\, \overline{\lambda}=-\,b\, \overline{\mu}=-\alpha.  $$

From this it follows that $|\frac{-\alpha}{a}|^2 + |\frac{\alpha}{b}|^2=1$, that is,
$|\alpha|^2= \frac{|a|^2 \|b|^2}{|a|^2 +|b|^2}< \frac{1}{4}.$

Note that if $a\,b=0$, then $\alpha=0$.

\qed

\begin{lemma} \label{Lem28} Given positive integers $d,D$, where $1<d$ and $D> 2 d +2$, denote
$$ f(t) = (1-t)^{d+1-D}\,(1+t)^{1-d}+ (1+t)^{d+1-D}\,(1-t)^{1-d}.$$

then, $f(t)$ is increasing on the interval $(0,1)$.

\end{lemma}

{\bf Proof:}

For any $t\in (0,1)$
we have

$$ f'(t)\,= (1-t)^{d+1-D}\,(1+t)^{1-d}\,[\frac{1-d}{1+t}- \frac{d+1-D}{1-t}]+$$
$$  (1+t)^{d+1-D}\,(1-t)^{1-d}\,[  \frac{d+1-D}{1+t}-\frac{1-d}{1-t}].$$

Taking $z=\frac{1+t}{1-t}>1$ we get
$$ (1+t)^{D-1}\, f'(t)= z^{D-d-1} \,[(D-d-1)\,z - (d-1)] + z^{d-1}\,[(d-1)\,z-(D-d-1)\,]>$$
$$ z^{ D-d-1}  [ \,(D-d-1)-\,(d-1)\,] + z^{d-1} \,[\,(d-1) - (D-d-1)\,] \,= $$
$$(z^{D-d-1} \,-z^{d-1})\, (D- 2\,d)>0,$$
because $z>1$,

\qed

\medskip

Suppose $0\leq a < 1/4$ and $d,D$ positive integers such that $1<d<D-1$. Define

$$J(d,D,a)\,=\frac{(D-1)\,!\, }{(d-1)\,!\, (D-d-1)\, ! } \int_{1/2 - \sqrt{1/4-a} }^{1/2 + \sqrt{1/4-a}} \frac{(w\, (1-w)- a)^{D-2} }{w^{D-d-1} \, (1-w)^{d-1}}  \, dw.$$

\medskip

\begin{lemma} \label{Lem29} Suppose $d,D $ are positive integers $1<d,\, 2\,d+2 < D$. Then,
$$0\leq J(d,D,a)< e^{- 4 \,a (D-3/2)},\,\,\,\,\text{where}\,\,\,0\leq a <1/4.$$

\end{lemma}

{\bf Proof:} Note that $J(d,D,a)$ is positive.

In the integration we divide the integral in two parts:  $[1/2 - \sqrt{1/4-a},\,1/2]$ and $[1/2,\,1/2 + \sqrt{1/4-a}]$.

We make a change of variable $w=  1/2\,-\sqrt{1/4-x}$ on the first interval and  $w=  1/2\,+\sqrt{1/4-x}$ on the second interval.
On both cases we get $x=w\,(1-w)$ and $a\leq x\leq 1/4.$

From this it follows

$$J(d,D,a)\,=$$
$$\frac{(D-1)\,!\, }{(d-1)\,!\, (D-d-1)\, ! } \int_{a}^{1/4} (x-a)^{D-2} [(\frac{1}{2}- \sqrt{\frac{1}{4}-x})^{d+1-D}\,(\frac{1}{2}+ \sqrt{\frac{1}{4}-x})^{1-d}+$$
$$ (\frac{1}{2}+ \sqrt{\frac{1}{4}-x})^{d+1-D}\,(\frac{1}{2}- \sqrt{\frac{1}{4}-x})^{1-d}\,\,] \frac{1}{2\, \sqrt{1/4-x}}\, dx=$$
$$\frac{2^{D-2}\,(D-1)\,!\, }{(d-1)\,!\, (D-d-1)\, ! } \int_{a}^{1/4} (x-a)^{D-2} [(1- \sqrt{1-4\,x})^{d+1-D}\,(1+ \sqrt{1-4\, x})^{1-d}+$$
$$(1+ \sqrt{1-4\,x})^{d+1-D}\,(1- \sqrt{1-4\, x})^{1-d}\, ]\,\frac{1}{\sqrt{1-4\,x}\,}\,dx.$$

Now we consider $y=\frac{x-a}{ 1/4 -a}$. In this case
$(1-4x)=(1-4a)(1-y)$.

Then,
$$ J(d,D,a) =$$
$$\frac{(1-4a)^{3/2}\,(D-1)\,!\, }{2^D\,(d-1)\,!\, (D-d-1)\, ! }\,\int_0^1\,y^{D-2} [\,(1-\sqrt{1-4a}\,\sqrt{1-y})^{d+1-D} $$
$$ (1+\sqrt{1-4a}\,\sqrt{1-y})^{1-d}+ (1+\sqrt{1-4a}\,\sqrt{1-y})^{d+1-D} $$
$$(1-\sqrt{1-4a}\,\sqrt{1-y})^{1-d}  \,\,]\, \frac{1}{\sqrt{1-y}}\, dy. $$

Note that just the expression under $[\,\,$ $\,\,]$ depends on $a$.
For each $y\in (0,1)$ we have $\sqrt{1-4a}\,\sqrt{1-y}\in(0,1)  $ is an decreasing function of $a$. It follows from Lemma \ref{Lem28} that for each $y\in (0,1)$ the integrand is a decreasing function of $a$.

Therefore, $\frac{J(d,D,a)}{(1-4a)^{D-3/ 2}}$ is a decreasing function of $a$.
As $J(d,D,0)=1$ (see Lemma \ref{Lem23}) it follows that
$$J(d,D,a)\leq (1- 4\, a)^{D-3/2},\,\,\,\,0 \leq a < 4.$$

Finally, note that $(1- 4\, a)^{D-3/2}\leq e^{ -4\, a\, (D-3/2) }$

\qed

\medskip

\begin{corollary} \label{Cor30}
 If $1<d, \,D>2\, d +2$ and $\frac{\log D}{D}<\frac{1}{3}$, then

$$J(d,D,a) < D^{-3} e^{ \frac{9\, \log D}{2\,D}},\,\,\,\,\text{where}\,\, a=\frac{3}{4}\, \frac{\log D}{D}.
$$

\end{corollary}

{\bf Proof:} It follows from Lemma \ref{Lem29} because $ 0<\frac{3}{4} \frac{\log D}{D} < \frac{1}{4}.$

\qed

\medskip

\begin{lemma} \label{Lem31}

Suppose $1\leq \nu\leq N $, $3<d_\nu$, $D>2 \, d_\nu+2$ and $ \frac{\log D}{D}<\frac{1}{5}.$

Then,
\begin{equation} \label{fin} \int_\Delta \max_{1\leq \alpha \neq \beta\leq D}\,|\,< \phi_\alpha ,P_\nu ( \mathcal{D})  \phi_\beta>\,|^2\, w_\Delta(\mathcal{D})< \frac{\log D}{D}\end{equation}
where $\phi_1,..,\phi_D$ is an orthonormal basis of eigenvectors for $H$ (without resonances).

\end{lemma}

{\bf Proof:} By Lemma \ref{Lem23} and Corollary \ref{Cor30} the probability that the integrand is bigger than $a$ is
smaller than
$$ \frac{D\, (D-1)}{2}\, D^{-3} e^{\frac{9\, \log D}{2\,D}},\,\,\,\,\,a=\frac{3\, \log D}{4\, D},$$
because as $e_{\alpha,\beta} = \overline{e_{\beta,\alpha}}$ we just have to take $\alpha<\beta$.

By the Remark before Lemma \ref{Lem24} the integral is smaller than
$$ \frac{3}{4}\, \frac{\log D}{D} \, + \frac{D\,(D-1)}{8} D^{-3}\, e^{\frac{9\,\log D}{2\, D}},$$
because by Lemma \ref{Lem27} $|e_{\alpha,\beta} |<1/4 $.

As $D-1<D$ we have
$$  \frac{\frac{D\,(D-1)}{8} D^{-3}\, e^{\frac{9\,\log D}{2\, D}}}{\frac{\log D}{D} }$$
$$  < \,\frac{1}{8\, D}  e^{\frac{9\,\log D}{2\, D}}\, \frac{D}{\log D}= e^{\frac{9\,\log D}{2\, D}}\, \frac{1}{8\,\log D}.$$

Now, as $D\geq 9$, $\log D\geq 2$, we get
$$ \frac{1}{8\, \log D} e^{\frac{9\,\log D}{2\, D}}< \frac{1}{16} e^{9/10} < \frac{e}{16}<1/4.$$

Now we put the two estimates together $\frac{3}{4}\, \frac{\log D}{D} + \frac{1}{4}\, \frac{\log D}{D}$  and we get the claim of the Lemma.

\qed

\medskip

The Lemma \ref{88} follows from Lemmas \ref{Lem26} and \ref{Lem31}.
In this way we get the claim of the Quantum  Ergodic Theorem of von Neumann.

\bigskip

\section{Appendix 1}

In this Appendix we will show that

\begin{equation} \label{xuxa}
 \frac{1}{\text{vol (S)}   } \int_S \, (\,\sum_{j=1}^d  |x_j|^2\,)^2\, d\, S(x)= \frac{d^2 + d}{D\, (D+1)}.
\end{equation}

First we will show  that when $S$ is the unitary sphere in $\mathbb{R}^n$, $m\geq 1$ and $n \geq m$, then

\begin{equation} \label{xuxu}
 \int_S \, (\,\sum_{j=1}^m  |x_j|^2\,)^2\, d\, S(x)=  \text{vol (S)} \frac{m^2 + 2 m}{n\, (n+2)}.
\end{equation}

It is easy to see that (\ref{xuxa}) follows from (\ref{xuxu}).

\medskip

1) $\int x_j^2  d\, S(x)= \frac{vol(S)}{n} $ for $j=1,2,...,n$ because the integral does not depend of $j$.

2) Suppose $B$ is the unitary ball in $\mathbb{R}^n$. Consider in polar coordinates
$$ T: S \times [0,1] \to B,$$
where $T(x,\rho)= \rho \,x.$

Then, $T^* (dx_1\wedge...\wedge d x_n)= \rho^{n-1} d S (x) \wedge d \rho$.

Therefore,
$$ \int_B (x_1^2 +..+x_n^2) d x_1...dx_n= \int_{S \times [0,1]} T^* ( (x_1^2 +..+x_n^2  ) dx_1\wedge...\wedge dx_n)=$$
$$\int_{S \times [0,1]}  \rho^{n+1} d S(x) \wedge d \rho = \text{vol}(S) \, \int_0^1 \rho^{n+1} d \rho = \frac{\text{vol} (S)}{n+2}.$$

Finally, $\int_B x_j ^2 dx_1...dx_n= \frac{\text{vol} (S)}{n\,(n+2)}$ because it is independent of $j=1,2...,n.$

3) For $j=1,2...,n$, we have
$$ \int_S x_j^4\, d S(x) = 3 \, \int_B x_j^2 \, dx_1...dx_n= \frac{3 \text{vol} (S)}{n\, (n+2)}$$
by the divergent theorem and by 2) above.

4) If $\leq i< j\leq n$, then
$$\int_S x_i^2\, x_j^2 \,d S(x) = \int_B x_j^2 \, dx_1...dx_n= \frac{ \text{vol} (S)}{n\, (n+2)}.$$
by the divergent theorem and by 2) above.

\medskip

The integral

$$  \int_S \, (\,\sum_{j=1}^d  |x_j|^2\,)^2\, d\, S(x)$$
is a sum of terms of the kind $\int_S x_i^2\, x_j^2 \,d S(x)$, $i \neq j$, and $\int_S x_j^4\,  \,d S(x)$, $j=1,2,..n$.

Just collecting the different terms and using the estimates above we get the initial claim (\ref{xuxu}).

\medskip

\section{Appendix 2 - proof of Lemma \ref{20}}

Suppose $\epsilon>0$ is small enough, consider
$$ f\,|_{f^{-1} (a-\epsilon,\, a+\epsilon)} :  f^{-1} (a-\epsilon,\, a+\epsilon)\to (a-\epsilon,\,a+ \epsilon).$$

Given $h\in \mathbb{R}$, $0<|h|<\epsilon$, then integrating $(g \circ f)\, \lambda$ we get

$$G(a+h)-G(a)=
\, \int_0^h g(a+t)\, dt\, \int_{X_{a+t}} \, \frac{ \lambda_{a+t}}{|\, \,\text{ grad}\, f\,|}.$$
(where $\lambda_v$ is the volume form on $X_v =f^{-1}(v)$ for $v \in (a-\epsilon,\,a+ \epsilon))$ because $df (\text{grad} f)=
|\,\text{ grad}\, f\,|^2$. From this follows that for some $0\leq \theta\leq 1$ we have
$$ G(a+h)-G(a)=
\, h \,g(a+\,\theta\,h)\, \, \int_{X_{a+\,\theta\,h}} \, \frac{ \lambda_{a+\,\theta\,h}}{|\, \,\text{ grad}\, f\,|},$$

Now we divide the above expression by $h$ and we take the limit when $h\to 0$

\qed

\bigskip

{\bf Instituto de Matematica - UFRGS - Brasil}
\medskip

A. O. Lopes was partially supported by CNPq and INCT.

\end{document}